\RequirePackage{snapshot}
\documentclass[longauth]{aa}
\newcommand{\colfigsize}{\hsize}

\usepackage{graphicx}

\usepackage{txfonts}
\usepackage[dvipsnames]{xcolor}
\definecolor{cadmiumgreen}{rgb}{0.0, 0.7, 0.2}
\usepackage{hyperref}
\hypersetup{
  breaklinks=true,
  colorlinks = true,
   citecolor = cadmiumgreen
}
\usepackage{multirow}

\usepackage{natbib,twoopt}
\bibpunct{(}{)}{;}{a}{}{,} 

\newcommand{\Rsun}{$R_\odot$\xspace}
\newcommand{\Msun}{$M_\odot$\xspace}

\newcommand{\Mearth}{$M_{\oplus}$\xspace}
\newcommand{\Rearth}{$R_{\oplus}$\xspace}

\newcommand{\fig}[1]{Fig.~\ref{#1}\xspace}
\newcommand{\Fig}[1]{Figure~\ref{#1}\xspace}

\newcommand{\cms}{%
    \ifmmode {\rm cm/s}%
    \else ${\rm cm\,s^{-1}}$\xspace%
    \fi%
}
\newcommand{\ms}{%
    \ifmmode {\rm m\,s^{-1}}%
    \else ${\rm m\,s^{-1}}$\xspace%
    \fi%
}
\newcommand{\kms}{%
    \ifmmode {\rm km/s}%
    \else ${\rm km/s}$\xspace%
    \fi%
}
\newcommand{\msy}{%
    \ifmmode {\rm m\,s^{-1}{\rm yr}^{-1}}%
    \else ${\rm m\,s^{-1}}{\rm yr}^{-1}$\xspace%
    \fi%
}
\newcommand{\msyy}{%
    \ifmmode {\rm m\,s^{-1}{\rm yr}^{-2}}%
    \else ${\rm m\,s^{-1}}{\rm yr}^{-2}$\xspace%
    \fi%
}

\newcommand{\vsys}{\ifmmode v_{\rm sys} \else $v_{\rm sys}$\xspace \fi}
\newcommand{\logrhk}{\ifmmode \log R'_{HK} \else $\log R'_{HK}$\xspace \fi}

\newcommand{\proxima}{Proxima\xspace}

\definecolor{linkcolor}{rgb}{0.1216,0.4667,0.7059}
\usepackage{xspace}

\makeatletter
\newcommand\cp[1]{%
   \@ifundefined{al@#1\@extra@b@citeb}{\citep{#1}\xspace}{\citepalias{#1}\xspace}%
}
\newcommand\ct[1]{%
   \@ifundefined{al@#1\@extra@b@citeb}{\citet{#1}\xspace}{\citetalias{#1}\xspace}%
}
\makeatother

\defcitealias{SuarezMascareno2020}{SM2020}

\begin{document}

   \title{A candidate short-period sub-Earth\\orbiting Proxima Centauri%
      \thanks{Based on Guaranteed Time Observations collected at the European
              Southern Observatory (ESO) by the ESPRESSO
              Consortium under ESO programmes 1102.C-0744,
              1102.C-0958, 1104.C-0350, and 106.21M2.}}

   \author{
      J.~P.~Faria \inst{\ref{ia-porto}, \ref{fcup}} \and
      A.~Suárez~Mascareño \inst{\ref{iac}, \ref{iacU}} \and
      P.~Figueira \inst{\ref{eso-santiago}, \ref{ia-porto}} \and
      A.~M.~Silva \inst{\ref{ia-porto}, \ref{fcup}} \and
      M.~Damasso \inst{\ref{inaf-torino}} \and
      O.~Demangeon \inst{\ref{ia-porto}, \ref{fcup}} \and
      F.~Pepe \inst{\ref{geneva}} \and
      N.~C.~Santos \inst{\ref{ia-porto}, \ref{fcup}} \and
      R.~Rebolo \inst{\ref{csic}, \ref{iacU}, \ref{iac}} \and
      S.~Cristiani \inst{\ref{inaf-trieste}} \and
      V.~Adibekyan \inst{\ref{ia-porto}} \and
      Y.~Alibert \inst{\ref{bern}} \and
      R.~Allart \inst{\ref{montreal}, \ref{geneva}} \and
      S.~C.~C.~Barros \inst{\ref{ia-porto}, \ref{fcup}} \and
      A.~Cabral \inst{\ref{ia-fcul}, \ref{fcul}} \and
      V.~D'Odorico \inst{\ref{inaf-trieste}, \ref{trieste1}, \ref{pisa}} \and
      P.~Di~Marcantonio \inst{\ref{inaf-trieste}} \and
      X.~Dumusque \inst{\ref{geneva}} \and
      D.~Ehrenreich \inst{\ref{geneva}} \and
      J.~I.~González~Hernández \inst{\ref{iacU}, \ref{iac}} \and
      N.~Hara \inst{\ref{geneva}} \and
      J.~Lillo-Box \inst{\ref{madrid-astrobio}} \and
      G.~Lo~Curto \inst{\ref{eso-garching}, \ref{eso-santiago}} \and
      C.~Lovis \inst{\ref{geneva}} \and
      C.~J.~A.~P.~Martins \inst{\ref{ia-porto}, \ref{caup}} \and
      D.~Mégevand \inst{\ref{geneva}} \and
      A.~Mehner \inst{\ref{eso-santiago}} \and
      G.~Micela \inst{\ref{inaf-palermo}} \and
      P.~Molaro \inst{\ref{inaf-trieste}, \ref{trieste1}} \and
      N.~J.~Nunes \inst{\ref{ia-fcul}} \and
      E.~Pallé \inst{\ref{iac}, \ref{iacU}} \and
      E.~Poretti \inst{\ref{inaf-brera}} \and
      S.~G.~Sousa \inst{\ref{ia-porto}, \ref{fcup}} \and
      A.~Sozzetti \inst{\ref{inaf-torino}} \and
      H.~Tabernero \inst{\ref{madrid}} \and
      S.~Udry \inst{\ref{geneva}} \and
      M.~R.~Zapatero~Osorio \inst{\ref{madrid}} }
   \institute{
      Instituto de Astrofísica e Ciências do Espaço, Universidade do Porto, CAUP, Rua das Estrelas, 4150-762 Porto, Portugal \label{ia-porto} \and
      Departamento de Física e Astronomia, Faculdade de Ciências, Universidade do Porto, Rua Campo Alegre, 4169-007 Porto, Portugal \label{fcup} \and
      Instituto de Astrofísica de Canarias, 38205 La Laguna, Tenerife, Spain \label{iac} \and
      Departamento de Astrofísica, Universidad de La Laguna, 38206 La Laguna, Tenerife, Spain \label{iacU} \and
      European Southern Observatory, Alonso de Cordova 3107, Vitacura, Santiago, Chile \label{eso-santiago} \and
      INAF - Osservatorio Astrofisico di Torino, Via Osservatorio 20, 10025 Pino Torinese, Italy \label{inaf-torino} \and
      Département d'astronomie de l'Université de Genève, Chemin Pegasi 51, 1290 Versoix, Switzerland \label{geneva} \and
      Consejo Superior de Investigaciones Científicas, 28006 Madrid, Spain \label{csic} \and
      INAF - Osservatorio Astronomico di Trieste, Via Tiepolo 11, 34143 Trieste, Italy \label{inaf-trieste} \and
      Physics Institute of University of Bern, Gesellschaftsstrasse 6, 3012 Bern, Switzerland \label{bern} \and
      Department of Physics, and Institute for Research on Exoplanets, Université de Montréal, Montréal, H3T 1J4, Canada \label{montreal} \and
      Instituto de Astrofísica e Ciências do Espaço, Faculdade de Ciências da Universidade de Lisboa, Edifício C8, Campo Grande, 1749-016, Lisbon, Portugal \label{ia-fcul} \and
      Faculdade de Ciências da Universidade de Lisboa (Departamento de Física), Edifício C8, 1749-016 Lisboa, Portugal \label{fcul} \and
      Institute for Fundamental Physics of the Universe, IFPU, Via Beirut 2, 34151 Grignano, Trieste, Italy \label{trieste1} \and
      Scuola Normale Superiore, Piazza dei Cavalieri, 7 I-56126 Pisa, Italy \label{pisa} \and
      Centro de Astrobiología (CAB, CSIC-INTA), Depto. de Astrofísica, ESAC campus, 28692, Villanueva de la Cañada (Madrid),\!Spain \label{madrid-astrobio} \and
      European Southern Observatory, Karl-Schwarzschild-Str. 2, D-85748 Garching bei München, Germany \label{eso-garching} \and
      Centro de Astrofísica da Universidade do Porto, Rua das Estrelas, 4150-762 Porto, Portugal \label{caup} \and
      INAF - Osservatorio Astronomico di Palermo, Piazza del Parlamento 1, 90134 Palermo, Italy \label{inaf-palermo} \and
      INAF - Osservatorio Astronomico di Brera, Via E. Bianchi 46, 23807 Merate, Italy \label{inaf-brera} \and
      Centro de Astrobiología (CSIC-INTA), Crta. Ajalvir km 4, E-28850 Torrejón de Ardoz, Madrid, Spain \label{madrid} 
    }

   \date{Received 30 September, 2021; accepted 24 December, 2021}

  \abstract
   {Proxima Centauri is the closest star to the Sun. This small, low-mass, mid M
    dwarf is known to host an Earth-mass exoplanet with an orbital period of
    11.2 days within the habitable zone, as well as a long-period planet
    candidate with an orbital period of close to 5 years.}
   {We report on the analysis of a large set of observations taken with the
    ESPRESSO spectrograph at the VLT aimed at a thorough evaluation of the
    presence of a third low-mass planetary companion, which started emerging
    during a previous campaign.}
   {Radial velocities (RVs) were calculated using both a cross-correlation
    function (CCF) and a template matching approach. The RV analysis includes a
    component to model Proxima's activity using a Gaussian process (GP). We use
    the CCF's full width at half maximum to help constrain the GP, and we study
    other simultaneous observables as activity indicators in order to assess the
    nature of any potential RV signals.}
   {We detect a signal at $5.12\pm0.04$ days with a semi-amplitude of $39\pm7$
    \cms. The analysis of subsets of the ESPRESSO data, the activity indicators,
    and chromatic RVs suggest that this signal is not caused by stellar
    variability but instead by a planetary companion with a minimum mass of
    $0.26\pm0.05$ \Mearth (about twice the mass of Mars) orbiting at 0.029 au
    from the star. The orbital eccentricity is well constrained and compatible
    with a circular orbit.}
   {}


   \maketitle
%

\section{Introduction}

   The last couple of decades have seen a fast increase in the number of known
   exoplanets orbiting solar-type stars. Several dedicated space missions 
   (e.g. CoRoT: \citealt{Baglin2006}, Kepler: \citealt{Borucki2016}, TESS:
   \citealt{Ricker2010}, CHEOPS: \citealt{Broeg2013}) 
   and ground-based instruments (e.g. HARPS: \citealt{Mayor2003}, CARMENES:
   \citealt{Quirrenbach2010}, HARPS-N: \citealt{Cosentino2012a}, HPF:
   \citealt{Mahadevan2012} and soon NIRPS: \citealt{Wildi2017} and SPIRou:
   \citealt{Artigau2014}) 
   are or will soon be pushing the detection limits towards smaller and
   lower-mass planets, and they already allow for the detailed study of the
   atmospheres and internal compositions of individual planets
   \citep[e.g.][]{Ehrenreich2020}. In this context, M dwarfs continue to be
   prime targets for exoplanet searches due to their smaller sizes and masses,
   resulting in higher-amplitude transit and radial velocity (RV) signals from
   the planets.

   Coincidentally, several of the M dwarfs closest to the Sun are known or
   suspected to host a planetary system \citep[e.g.][]{Ribas2018, Bonfils2018,
   Jeffers2020, Diaz2019, Tuomi2019, Lillo-Box2020}. In particular, the
   discovery of an Earth-mass planet orbiting Proxima Centauri
   \citep{Anglada-Escude2016}, our closest stellar neighbour, was one of the
   most significant results in the field, in part because the planet orbits
   inside the habitable zone (HZ) of the star \citep[e.g.][]{Kopparapu2013}.
   More recently, a second candidate super-Earth was detected at a longer
   orbital period of 5.21 years \citep{Damasso2020a} but has so far eluded a
   clear detection with direct imaging or astrometry
   \citep{Gratton2020,Kervella2020,Benedict2020}

   The detection of Proxima b and similar planets orbiting M dwarfs demonstrates
   that high-precision RV measurements allow us to find Earth-mass planets that
   orbit inside, or close to, the HZ of their host stars. Unfortunately, several
   photospheric and chromospheric phenomena associated with the presence of
   active regions at the stellar surface induce RV variations that can mimic or
   contaminate a planetary signal \citep[e.g.][and references
   therein]{Fischer2016}. For M dwarfs in particular, such activity-induced RV
   variations can reach a few \ms, even for quiet stars
   \citep[e.g.][]{SuarezMascareno2017}. As such, the detection of Earth-mass
   planets imposes a detailed characterisation of stellar activity, and it has
   motivated the development of precise instrumentation, such as the
   state-of-the-art ESPRESSO spectrograph \citep{Pepe2021}, which combines the
   large collecting power of the Very Large Telescope (VLT) with a very high
   level of stability in order to achieve unprecedented RV precision.

   One of the first results from ESPRESSO was the independent confirmation of
   Proxima b \citep[][hereafter SM2020]{SuarezMascareno2020}, which also
   revealed the first hints of a shorter-period signal that could be caused by a
   low-mass planet. Here we report on follow-up observations that confirm the
   presence of this low amplitude signal, which is probably caused by a planet
   with a minimum mass of only 0.26 \Mearth and an orbital period of 5.12 days.
   We first describe the observations (Sect. \ref{sec:observations}) and the
   methods used to derive (Sect. \ref{sec:rv-determination}) and analyse (Sect.
   \ref{sec:rv-analysis}) the RVs. In Sect. \ref{sec:planet-nature} we assess
   the evidence for the planetary nature of the signal, and we discuss our
   results in Sect. \ref{sec:discussion}.

   \begin{table}
      \renewcommand{\arraystretch}{1.2}
      \centering
      \caption{Stellar properties of \proxima, compiled from the literature.}
      \label{tab:parameters}
      \begin{tabular}[center]{l l l}
      \hline
      Parameter & Value & ref. \\ \hline\hline
      Parallax [\emph{mas}]      &  768.50 $\pm$ 0.20 & 1 \\
      Distance [pc]              & 1.3012 $\pm$ 0.0003 & 1 \\
      $m_{V}$  [mag]             & 11.13 $\pm$ 0.01 & 2 \\
      Spectral type              & M5.5V & 3 \\
      $L_{*}$ / $L_{\odot}$      & 0.0016 $\pm$ 0.0006 & 4 \\
      $T_{\rm eff}$ [K]          & 2900 $\pm$ 100 & 5 \\
      $M_*$ [\Msun\!]            & 0.1221   $\pm$ 0.0022 & 6 \\
      $R_*$ [\Rsun\!]            & 0.141 $\pm$ 0.021 & 4 \\
      $P_{\rm rot}$ [days]       & 90 $\pm$ 4 & 7 \\
      HZ range [au]              & 0.0423 -- 0.0816 & 8\\
      HZ periods [days]          & 9.1 -- 24.5 & 8 \\
      \hline
      \end{tabular}
      \tablefoot{The references for each parameter are as follows: 
      1 - \cite{GaiaCollaboration2016}
      2 - \citet{Jao2014}
      3 - \citet{Bessell1991}, 
      4 - \citet{Boyajian2012}, 
      5 - \citet{Pavlenko2017}, 
      6 - \citet{Mann2015}, 
      7 - \citet{Klein2021},
      8 - \citet{Delfosse2000}.
      These values match those used by \ct{SuarezMascareno2020}, although we
      only use the stellar mass determination from \citet{Mann2015}.}
   \end{table}

\section{ESPRESSO observations}
\label{sec:observations}

   \proxima (see Table \ref{tab:parameters} for the stellar parameters) is on
   the target list of the ESPRESSO Guaranteed Time Observations (GTO) survey,
   which is monitoring a select group of nearby stars with the goal of
   discovering low-mass planets in the HZ \citep{Hojjatpanah2019, Pepe2021}. The
   first planet discovery from the survey was presented very recently in
   \citet{Lillo-Box2021}.
   After an initial campaign aimed at confirming \proxima~b
   \cp{SuarezMascareno2020}, we restarted observations of \proxima in order to
   confirm a candidate signal close to 5 days. In this new campaign, the
   observational strategy was tailored to this period, with a typical interval
   of 1-2 days between observations and two exposures per night when possible. 

   To add to the 67 observations reported in \ct{SuarezMascareno2020}, we
   obtained 52 new ESPRESSO spectra of \proxima, for a total of 117 observations
   spread over 99 individual nights from 2019-02-10 to 2021-05-06. The
   measurements were taken in ESPRESSO's high resolution mode (HR21) with an
   exposure time of 900 seconds%
   \footnote{%
   With the exception of six observations, for which the exposure times were
   increased to compensate for bad atmospheric conditions.
   }. %
   We used the Fabry P\'erot (FP) for simultaneous calibration, which allows the
   instrumental drift to be monitored with a precision better than 10 \cms
   \citep{Wildi2010}. 
   
   After a careful analysis of each spectrum, we decided to exclude three
   observations that were affected by instrumental issues. On the night of 25
   April 2019, there was a cooling water temperature increase, which propagated
   inside the spectrograph. The drift measured on the red detector was 0.007
   pixel (i.e. 3 \ms). This spectrum also shows a signal-to-noise ratio below 1.
   On the observation of 31 July 2019, the FP exposure was saturated, most
   likely due to an issue with the neutral density filters. In this situation,
   the contamination by the FP light on the charge-coupled device may be large,
   and the drift measurement is likely unreliable. Finally, on the night of 10
   April 2021, the reported temperature of the atmospheric dispersion
   compensator (ADC) was 0\,K, revealing an issue with the correction.
   The barycentric Julian days of these three excluded spectra are
   2458598.523839, 2458695.52992, and 2459314.583084.

   In June 2019, ESPRESSO underwent an intervention to update the fibre link,
   which improved the instrument's efficiency by up to 50\% \cp{Pepe2021}. This
   intervention introduced an RV offset, leading us to consider separate
   ESPRESSO18 and ESPRESSO19 datasets%
   \footnote{But we note the slight misnomer, as both subsets actually only
   contain observations done in 2019.}.
   More recently, operations at Paranal were interrupted due to the COVID-19
   pandemic and ESPRESSO was taken out of operations between 24 March 2020 and
   24 December 2020. This led to a large gap in the observations after the
   initial campaign. Moreover, a change in one of the calibration lamps after
   the ramp-up of the instrument is likely to have introduced another RV offset.
   Therefore, we consider an independent ESPRESSO21 dataset for data obtained
   after the ramp-up.
   In summary, we have 114 available RVs, divided between ESPRESSO18 (50
   points), ESPRESSO19 (15 points), and ESPRESSO21 (49 points) subsets. The full
   time span of the data is 815 days, which we note is about 2.3 times shorter
   than the orbital period of \proxima c \citep{Damasso2020a}.
   
\section{Radial velocity determination}
\label{sec:rv-determination}

   ESPRESSO data were reduced with the instrument's data reduction software
   (DRS), version 2.2.8 \cp{Pepe2021}. The DRS transforms the raw images into
   wavelength-calibrated 2D spectra (in order-pixel space) through a series of
   recipes, which includes bias and dark subtraction, optimal extraction of
   spectral orders, flat-fielding and de-blazing, wavelength calibration, and
   computation of the barycentric correction and instrumental drift.

   From the extracted spectra, we used two techniques to determine the RVs, one
   based on the cross-correlation function (CCF) and the other on a
   template-matching (TM) algorithm, described below.

   \subsection{CCF}
      
      The ESPRESSO DRS automatically derives the RV of each extracted spectrum
      using the cross-correlation technique
      \citep[e.g.][]{Baranne1996,Bouchy2001}. In this case, the mask used to
      build the CCF was derived from an ESPRESSO spectrum of \proxima itself,
      with a line-search algorithm based on the derivative of the spectrum. Each
      spectral order is cross-correlated individually, producing one CCF per
      order, which are then combined into a final CCF per spectrum.

      To trace the star's activity, we extract a number of activity indicators
      from the ESPRESSO spectra or from the CCF. The DRS calculates the CCF's
      full width at half maximum (FWHM), contrast (i.e. the relative depth of
      the CCF), and a shape indicator called CCF asymmetry
      \citep[see][]{Pepe2021}. Using \texttt{actin} \cp{GomesdaSilva2018}, we
      also measure activity indices based on the CaII H\&K, HeI, H$\alpha$, and
      NaI lines.

   \begin{figure*}
   \includegraphics[width=\hsize]{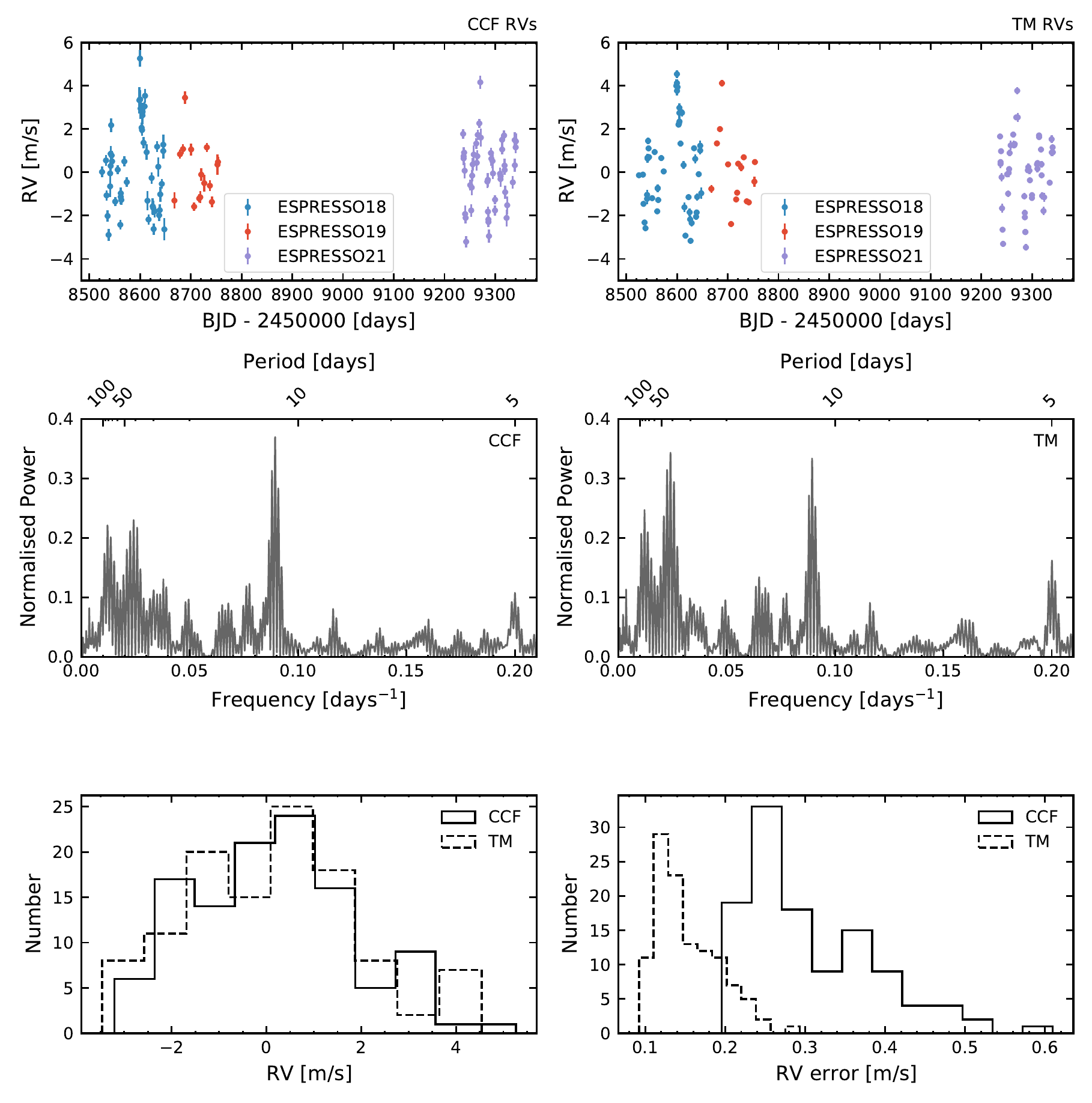}
   \caption{Comparison between RVs derived with the CCF and TM techniques. The
      top panels show the two RV time series, and the middle-left and
      middle-right panels show the generalised Lomb-Scargle (GLS) periodograms
      of CCF and TM RVs, respectively. In the bottom panels we display the
      histograms of the RVs (left) and of the RV uncertainties (right) for the
      two time series.}
   \label{fig:plot_DRS_TM}
   \end{figure*}

   \subsection{Template matching}

      We also consider a TM algorithm for RV extraction where each spectra is
      compared to a template spectrum built from the available ESPRESSO
      observations. This technique has been shown to improve RV precision when
      compared with the CCF method, in particular for M dwarfs
      \citep[e.g.][]{Anglada-Escude2012, Zechmeister2018, Lafarga2020}. A
      detailed description of the method may be found in \ct{Silva2021}. We
      provide here a brief overview.

      As a first step, a telluric template is created from a synthetic spectrum
      of Earth's transmittance, built with the TAPAS web interface
      \citep{Bertaux2014}. All regions where the absorption drops by more than
      1\% from the continuum (determined with a median filter) are flagged as
      telluric features, to be masked from the observed spectra. 

      To build a high signal-to-noise stellar template, all the observed spectra
      (as extracted by the DRS and before blaze correction) are combined order
      by order. The individual observations are RV shifted to a common reference
      frame, arbitrarily chosen to be that of the observation with the lowest
      uncertainty on the CCF RV, using the RVs derived previously with the CCF
      technique. The flux of each order is interpolated to a common wavelength
      grid and the mean flux of all observations is then calculated. We
      calculate the mean, instead of the sum, to avoid possible issues with
      numerical overflow. Uncertainties in the stellar template, originating
      from both the photon-noise of the individual spectra and from
      interpolation, are propagated \citep[see][for details]{Silva2021}.
      Importantly, we determine different stellar templates for each set of
      observations from ESPRESSO18, ESPRESSO19, and ESPRESSO21 to prevent any
      bias from instrumental systematics.

      Finally, and unlike other TM approaches, we use a single RV shift to
      describe simultaneously the differences between all orders of a given
      spectrum and the template. Within a Bayesian framework, we estimate the
      posterior distribution for the (single) RV shift, after marginalising with
      respect to a linear model for the continuum levels of the spectra and
      template. For computational tractability, we use a Laplace approximation
      \citep[see e.g.][chapter 8]{Bleistein1986} to characterise the posterior
      distributions for the RV shifts.

      From the Laplace approximation to the posterior distribution we use the
      mean as the estimated RV and the standard deviation as the estimated RV
      uncertainty, for each epoch. These uncertainties include the contributions
      from the photon noise of each observation, the interpolation used in
      constructing the template, as well as the order-to-order RV scatter. 

      For \proxima, this TM algorithm provides RVs that have lower uncertainties
      than those derived from the CCFs. \Fig{fig:plot_DRS_TM} shows a comparison
      of the two RV time series (top panels), highlighting the similar RV
      dispersion (bottom panel, left) but quite different distribution of RV
      uncertainties (bottom panel, right). The Lomb-Scargle periodograms
      \citep{Lomb1976,Scargle1982} of the two sets of RVs (middle panels), which
      have been calculated after subtraction of the weighted mean of each
      subset, also show differences. Most noticeably, the TM RVs present more
      power close to 5 days and close to the first harmonic of the stellar
      rotation period ($\sim$40 days). We also note the correlation coefficients
      between the two sets of RVs: $r=0.96$ (Pearson's) and $\rho = 0.95$
      (Spearman's).

      In summary, the two sets of RVs are compatible but the TM RVs are more
      precise. Throughout the paper, we perform the exact same analyses
      and describe the results for both the CCF RVs and the TM RVs.

\section{Radial velocity analysis}
\label{sec:rv-analysis}

   \begin{figure}
   \includegraphics[width=\colfigsize]{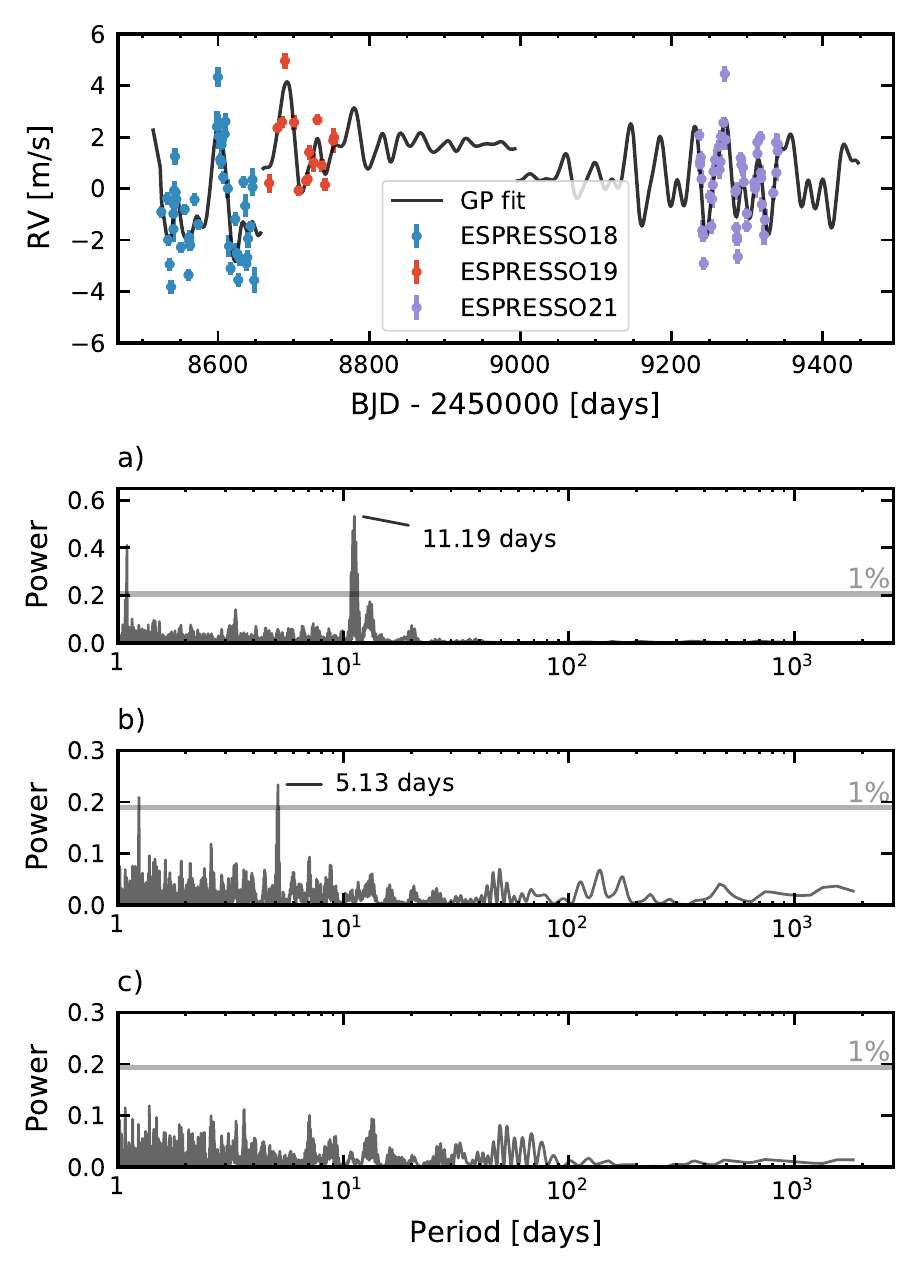}
   \caption{Pre-whitening procedure applied to the CCF RVs. The top panel shows
   the observed RVs together with the GP prediction. The periodogram of the
   residuals from this fit is shown in panel a), and the periodograms after two
   successive sinusoidal fits are in panels b) and c). The false alarm
   probability of 1\%, calculated with bootstrap randomisation, is shown by the
   horizontal grey lines.}
   \label{fig:pre_whitening}
   \end{figure}

   \subsection{Simple periodogram analysis}
   \label{sec:pre-white}

   We first consider a very simple pre-whitening procedure applied to the CCF
   RVs in order to determine the main periodicities present in the data. The
   observed RVs are initially modelled with a Gaussian process (GP) using a
   standard quasi-periodic (QP) kernel \citep{Rasmussen2006,Haywood2014}. For
   simplicity, the hyperparameters of the kernel are fixed to the values
   determined in \ct{SuarezMascareno2020} from a combined fit (their Table 3).
   Two RV offsets are adjusted between the three data subsets. The resulting fit
   is shown in the top panel of \fig{fig:pre_whitening} and the periodogram of
   the residuals is shown in panel a) of the same figure. The highest peak in
   the periodogram is at 11.19 days with a false alarm probability well below
   1\%.
   
   A sinusoidal signal is then fit to the residual RVs, with a period equal to
   that of the periodogram peak. The periodogram of the new residuals from this
   fit is shown in panel b), with a significant peak now appearing close to 5
   days and another peak (also significant) at an alias close to 1.2 days. After
   one additional sinusoidal fit to the residuals, no significant peaks are
   found in the periodogram (panel c).

   The two sinusoids recovered with this sequence of steps have semi-amplitudes
   equal to 85 \cms and 38 \cms, respectively, and the final residuals show an
   rms close to 50 \cms. Apart from the stellar activity signals that are being
   modelled by the initial GP, the 11-day and 5-day signals are the two main
   signals present in the RV data. However, this simple procedure is not
   sufficient to conclude on the planetary nature of the two signals. Moreover,
   the GP may be absorbing signals that are not caused by stellar activity and
   biasing the results. In the following section we build a more robust model
   considering Keplerian signals and perform a simultaneous analysis.

   \subsection{Joint RV and FWHM model}

   For a more robust determination of the model parameters, we analyse the RV
   and FWHM time series jointly, using a shared model for stellar activity. The
   FWHM was found to be an excellent proxy for the stellar activity of Proxima,
   showing clear variations at the stellar rotation period (80-90 days), and
   closely tracing the photometric behaviour of the star
   \cp{SuarezMascareno2020}. The same FWHM time series will be used for the
   analysis of RVs derived from the CCFs and from the TM approach.

   Our stellar activity model includes a GP with most of the hyperparameters
   shared between RVs and FWHM, but distinct variances. We tested two covariance
   functions to model stellar activity variations, the QP kernel mentioned above
   and the quasi-periodic with cosine (QPC) kernel proposed by
   \citet{Perger2021}. The results described below correspond to the QP kernel.
   Other assumptions for the activity model provide compatible results and are
   discussed briefly in Appendix \ref{app:covariances}.

   To model planetary RV signals, we assume Keplerian orbits parameterised by
   the orbital period $P$, semi-amplitude $K$, eccentricity $e$, mean anomaly
   $M_0$ at the time of the first observation, and argument of pericentre
   $\omega$. As a simple (and fast) criterion for dynamical stability, we only
   accept proposed solutions from the prior that are stable according to the
   angular momentum deficit (AMD) criterion \citep{Laskar2017}, using the
   extension proposed by \citet{Petit2017}, which takes mean-motion
   resonances into account.

   The subsets of RVs from ESPRESSO18, ESPRESSO19, and ESPRESSO21 are each
   assigned an independent RV offset and an additional jitter in the form of
   uncorrelated (white) noise characterised by its variance, and the same for
   the FWHM. A quadratic trend is also included in the RVs.
   The full model has a total of $4 + 5 + 2 N_s + 2(N_s-1) + 5 N_p$ parameters,
   where $N_s=3$ is the number of data subsets and $N_p$ is the number of
   Keplerians in the model. The different terms correspond to the average RV and
   FWHM plus the coefficients of the quadratic trend, the hyperparameters of the
   GP, the jitters, the instrument offsets, and finally the orbital parameters.
   
   Since we ultimately aim at comparing models with a different number of
   planets, it is important to carefully choose the prior distributions and that
   they encode prior knowledge independently of the value of $N_p$. 
   The GP amplitudes for the RVs and the FWHM ($\eta_1$) were assigned modified
   log uniform priors \citep[e.g.][]{Gregory2005} up to the span of each dataset
   (8.8 \ms for the RVs and 36.4 \ms for the FWHM) and with a break point at 1
   \ms.
   For the evolution timescale, $\eta_2$, we use a log uniform prior between 60
   and 400 days and for $\eta_3$, the parameter associated with the stellar
   rotation period, we use a uniform prior between 60 and 100 days. Proxima's
   rotation period was estimated to be 89.8 days (independently from RV data) to
   a precision of 4 days \citep[][see also
   \citealt{SuarezMascareno2016}]{Klein2021}; thus, this prior seems appropriate,
   and likely conservative.

   The same priors are used for the orbital parameters of all the $N_p$
   Keplerians in the model. We assign a wide log-uniform prior to the orbital
   periods, between 1 day and the full time span of observations (815 days), and
   a modified log-uniform prior for the semi-amplitudes, up to 10 \ms with a
   break point at 1 \ms. The eccentricities were assigned a Kumaraswamy prior
   \cp{Kumaraswamy1980}, which closely resembles the Beta distribution proposed
   by \citet{Kipping2013} but is easier to manipulate numerically. The two
   angular parameters, $M_0$ and $\omega$, were assigned uniform priors between
   0 and 2$\pi$. Other parameters such as offsets and jitters were assigned
   data-dependent but uninformative priors. The full set of priors is listed in
   Table \ref{tab:priors} and discussed further in Appendix \ref{app:priors}.

   To sample from the posterior distribution, we use the diffusive nested
   sampling (DNS) algorithm from \citet{Brewer2011}, as implemented in
   \texttt{kima} \cp{Faria2018}. Together with posterior samples, DNS provides
   an estimate for the marginal likelihood, or evidence, of the model, which we
   can use for model comparison \citep[e.g.][]{Brewer2014,Feroz2011}. 
   We obtain at least 50\,000 effective samples from the posterior -- as measured
   by the effective sample size (ESS), the number of samples with significant
   posterior weight -- for each model, which is more than enough to accurately
   characterise it.

   \subsection{Results from the joint model}

   \begin{figure}
      \centering
      \includegraphics[width=\colfigsize]{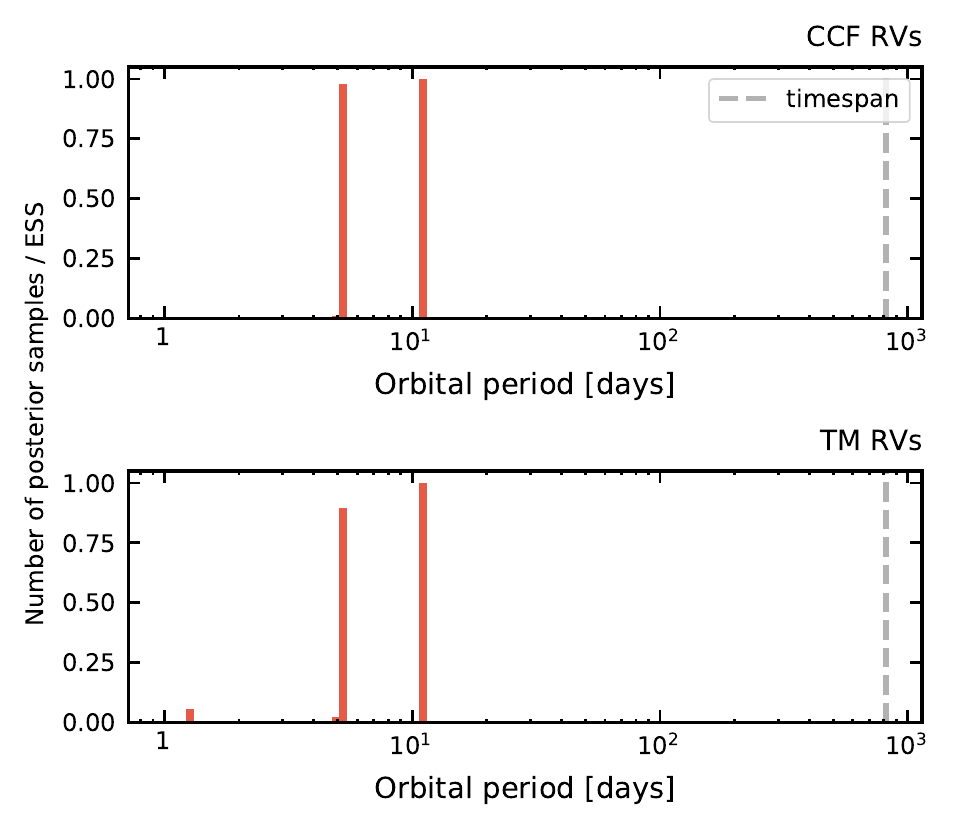}
      \caption{Posterior distribution for the orbital periods in the two-planet
      model from the analysis of the CCF RVs (top) and TM RVs (bottom). The
      posteriors are shown normalised by the ESS per
      histogram bin (so the maximum possible value in the abscissa is 1). The
      prior is log-uniform from 1 day to the time span of the data, which is
      marked with a dashed line.}
      \label{fig:posterior_period}
   \end{figure}

   \begin{figure*}
      \centering
      \includegraphics[width=0.49\hsize]{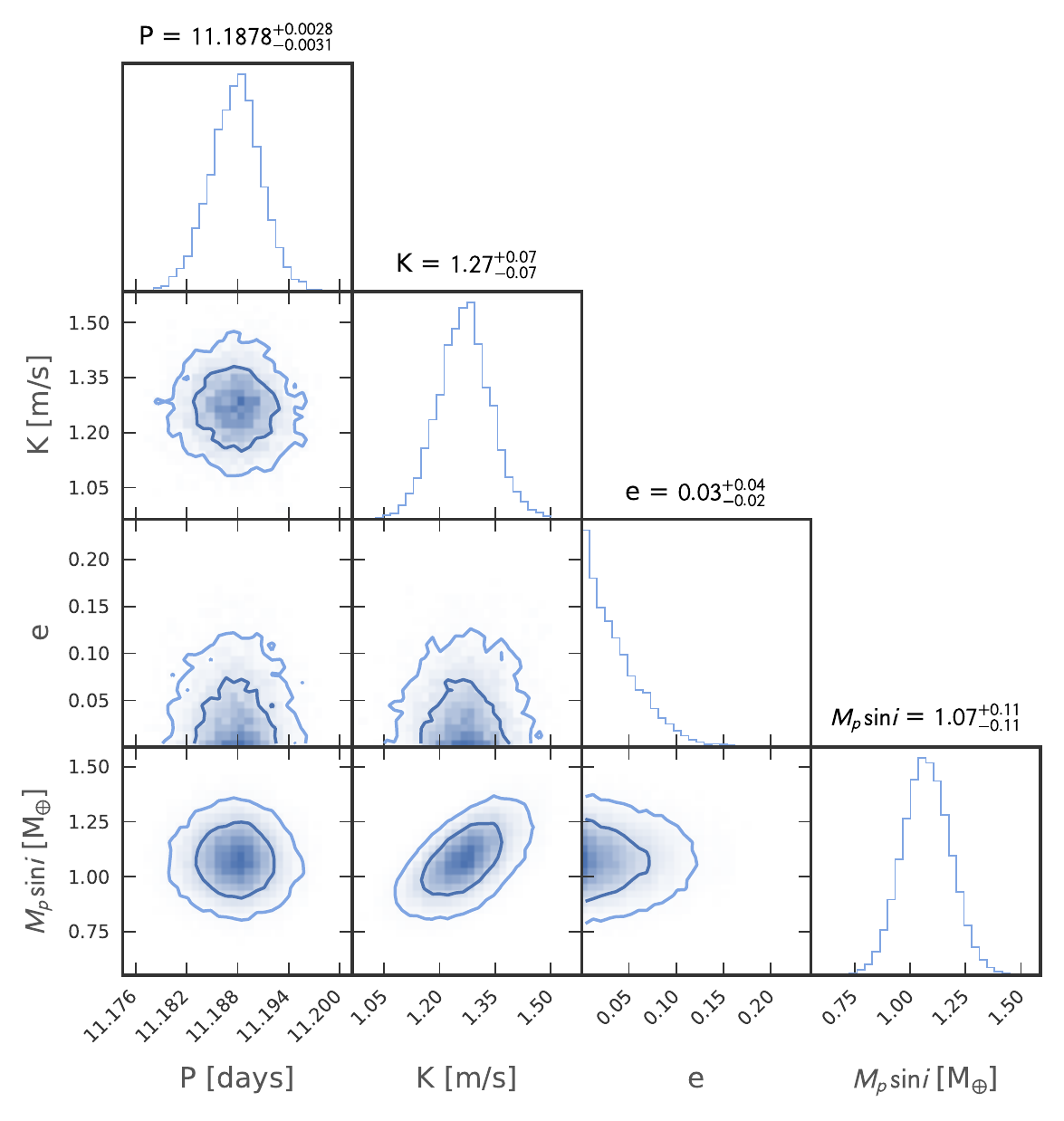}
      \includegraphics[width=0.49\hsize]{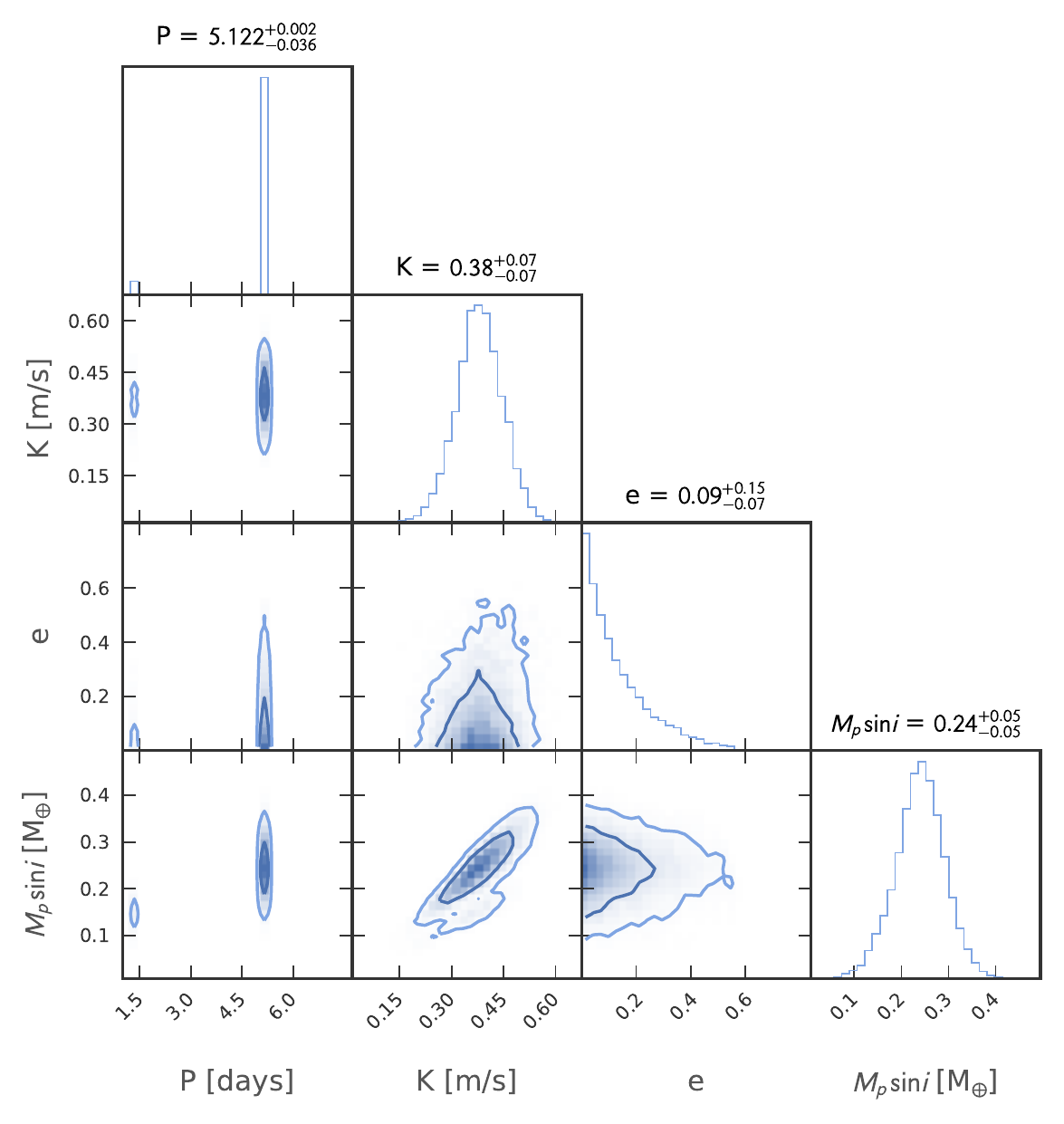}
      \caption{Joint and marginal posteriors for the orbital periods,
            semi-amplitudes, eccentricities, and minimum planet masses,
            $M_p\!\sin\!i$, in Earth masses, of the 11-day (left) and 5-day
            (right) Keplerian signals. These results come from the analysis of
            the TM RVs. The estimates at the top of each panel correspond to the
            median and 68\% quantiles of the posteriors.}
      \label{fig:corners_planets_TM}
   \end{figure*}

   The evidence values for models with $N_p = 0, 1, 2, 3$ are shown in Table
   \ref{tab:evidences}, together with the log Bayes factors between consecutive
   models. In the analysis of the CCF RVs, the model with $N_p=2$ is
   significantly preferred, with a $\Delta\ln Z > 5$, which corresponds to
   decisive evidence in the scale of \citet{Kass1995}.

   \begin{table}
      \caption{Evidence ($\ln Z$) and Bayes factors ($\Delta\ln Z$) for models
      with a given number of Keplerians, $N_p$, from the analysis of the CCF and
      TM RVs. The Bayes factors are calculated between models with $N_p$ and
      $N_p-1$ Keplerians. The evidence for the model assuming two circular
      orbits is also shown.}
      \label{tab:evidences}
      \centering
      \begin{tabular}{c c c c c}
         \hline
               & \multicolumn{2}{c}{CCF RVs} & \multicolumn{2}{c}{TM RVs} \\
         $N_p$ & $\ln Z$ & $\Delta\ln Z$ & $\ln Z$ & $\Delta\ln Z$ \\
         \hline\\[-0.5em]
         0 & -448.8 && -438.9 & \\
         1 & -414.0 & +34.8 & -389.5 & +49.4 \\
         2 & -408.9 & +5.1 & -385.6 & +3.9 \\
         3 & -410.2 & -1.3 & -387.3 & -1.7 \\
         \hline \\
         2 (circular) & -407.9 && -382.3 & \\
         \hline
      \end{tabular}
   \end{table}

   From this model, the posterior distribution for the orbital periods is shown
   in the top panel of \fig{fig:posterior_period}, where the count in each
   period bin was normalised by the ESS (thus resembling the TIP proposed by
   \citealt{Hara2021}). The posterior shows two very clear peaks at 5.12 days
   and 11.19 days. There is residual posterior probability close to 1.2 days,
   which is a 1-day alias of the 5.12-day period and an even smaller posterior
   peak at 43 days (both are not visible at the scale of the figure). The two
   main posterior peaks correspond to the orbital periods of Proxima b and of a
   candidate planet that we call Proxima d.

   While the period and semi-amplitude of the 5-day signal are well constrained,
   the posterior for the eccentricity is quite wide, with a peak at 0.45 (see
   \fig{fig:corners_planets_CCF}). Above this value, it also shows a sharp upper
   tail due to the AMD stability criterion, which makes higher eccentricities
   very improbable. The median and 68\% credible intervals result in an estimate
   for the eccentricity of $0.33^{+0.13}_{-0.23}$. 

   In this analysis of the CCF RVs, the 5-day signal has a semi-amplitude of 59
   \cms (posterior median), leading to a minimum mass of only $0.36 \pm 0.06$
   \Mearth, and a semi-major axis of 0.028 au, assuming it is due to a planetary
   companion. For both the 5-day and the 11-day signals, circular orbits are not
   excluded at the 2$\sigma$ level, which led us to consider a model where the
   eccentricities of both planets are fixed to zero. The evidence for this model
   (with both CCF and TM RVs) is also shown in Table \ref{tab:evidences}, and it
   is in both cases slightly higher than for the model with free eccentricities,
   although not significantly. This means that the currently available ESPRESSO
   data are not sufficient to distinguish between eccentric and circular orbits
   at the two-sigma level. 

   The analysis of the TM RVs provides qualitatively similar results, although
   the Bayes factor between the models with $N_p=1$ and $N_p=2$, even if still
   decisive, is not as high (Table \ref{tab:evidences}). The same two clear
   peaks at 5.12 days and 11.19 days are visible in the posterior for the
   orbital periods (shown in the bottom panel of \fig{fig:posterior_period}),
   but the alias at 1.2 days shows a higher posterior probability.
   
   In terms of likelihood alone, the solutions with the two orbital periods of
   5.12 days and 11.19 days are well isolated, with $\Delta \ln \mathcal{L} > 5$
   relative to any other combination of periods (except for the 1.2-day alias).
   The same is true for the analysis of the CCF RVs.
   Despite having comparable periodogram power in the analysis from Sect.
   \ref{sec:pre-white} (panel b in \fig{fig:pre_whitening}), the 5.12-day
   period shows a much larger posterior probability and likelihood when compared
   to the 1.2-day alias, in both RV datasets, leading to the conclusion that
   the former is the correct periodicity.
   
   In \fig{fig:corners_planets_TM}, we show the joint posterior distributions
   for the orbital period, semi-amplitude, eccentricity, and planet minimum mass
   of the two Keplerians, using the stellar mass in Table
   \ref{tab:parameters}. In the TM dataset, the semi-amplitude of the 5-day
   signal is significantly smaller than for the CCF RVs, at $38$ \cms
   (\fig{fig:corners_planets_TM}), although the results for the two datasets
   are compatible at less than 2$\sigma$. This ultimately leads to an estimate
   for the planetary mass that is 30\% lower than with the CCF RVs, at
   $0.24\pm0.05$ \Mearth (posterior median). The eccentricity is better
   constrained by the TM RVs and the solutions for both planets are compatible
   with circular orbits. We note that, for such a low semi-amplitude, the
   departure from a sinusoidal signal caused by an eccentricity of, say, 0.1,
   would be of the order of $K \cdot e \approx 5 \,\cms $, which is below the RV
   precision we can currently probe.

\begin{figure}
      \centering
      \includegraphics[width=\colfigsize]{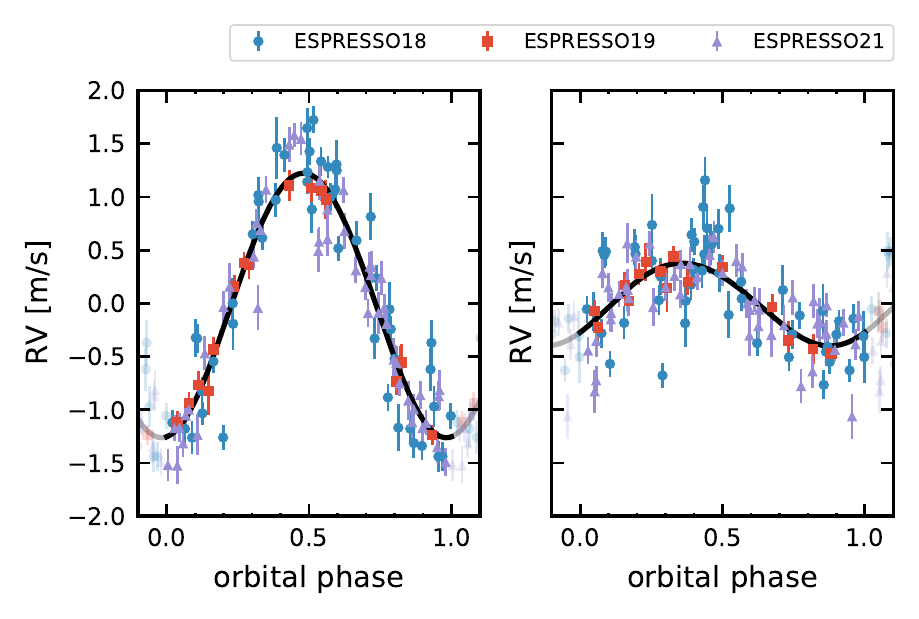}
      \caption{Template-matching RVs from ESPRESSO phase-folded to the orbital
       periods of \proxima b (left) and \proxima d (right). The GP and
       background components of the model were subtracted assuming the MAP
       solution. The full fit to the RVs and FWHM is shown in \fig{fig:2p_solution}.}
      \label{fig:phase_curve}
   \end{figure}

   \Fig{fig:phase_curve} shows the ESPRESSO RVs phase-folded to the orbital
   periods of the two Keplerians, assuming the maximum \emph{a posteriori} (MAP)
   solution from the two-planet model applied to the TM RVs. We finally adopt this
   solution and the 68\% quantiles of the posterior as the estimates for the
   parameters of the two-planet model, listed in Table \ref{tab:post} in Appendix
   \ref{app:posterior_estimates}. We note that the choice for quoting the
   results from the TM RVs over the CCF RVs is mostly guided by the
   substantially smaller RV uncertainties on this dataset.
   
   In \fig{fig:2p_solution}, we show the individual fits to the ESPRESSO18,
   ESPRESSO19, and ESPRESSO21 subsets as well as the combined residuals. The
   weighted RMS of the RV residuals reaches 27 \cms and 60 \cms for the FWHM. We
   note a smaller scatter in the ESPRESSO19 residuals (in both RV and
   FWHM), which could be explained by the fewer observations in this subset or
   the larger time separation between the points. Indeed, the
   ESPRESSO19 subset does not contain nights with more than one observation,
   which may suggest some degree of overfitting and an inability of the model to
   represent intra-night variations. In other words, the model may not
   completely capture all the RV (and FWHM) variations at short periods of 1 or
   2 days.

   \begin{figure}
      \centering
      \includegraphics[width=\colfigsize]{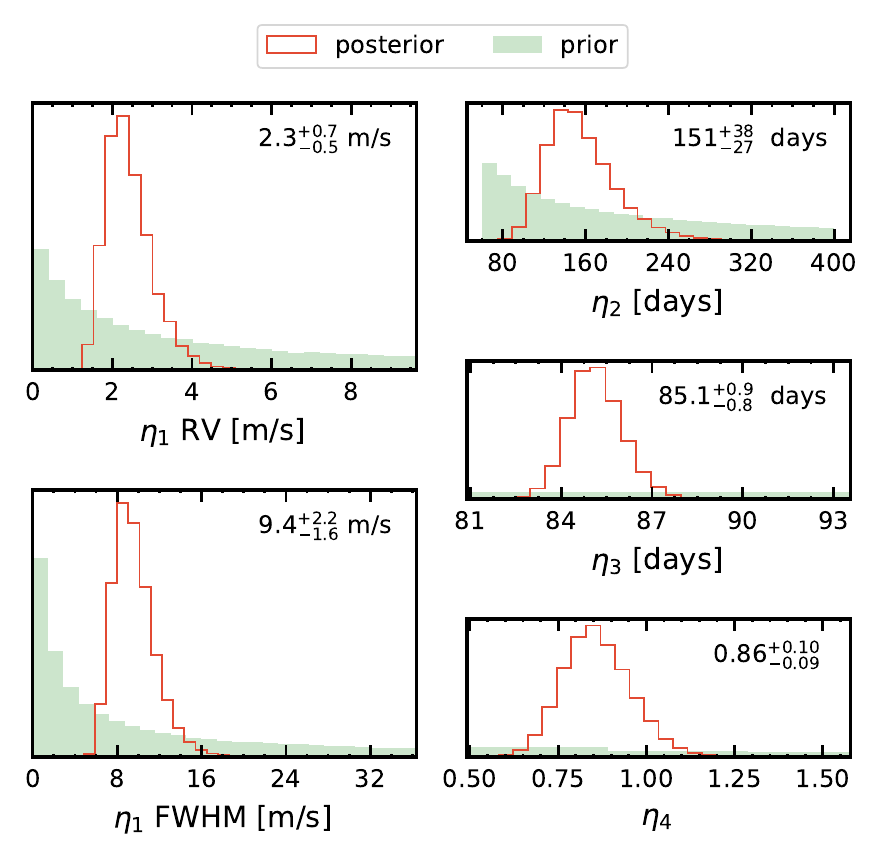}
      \caption{Prior and posterior distributions for the hyperparameters of the
               GP in the two-planet model. Posterior estimates (median and 68\%
               credible intervals) are show in each panel. It should be noted
               that for $\eta_3$ and $\eta_4$ the full support of the prior is
               larger than what is shown.}
      \label{fig:GP_parameters}
   \end{figure}

   The stellar activity component of the model (the shared GP for RVs and FWHM),
   is very well constrained both in the CCF and TM datasets. The posteriors for
   the GP hyperparameters from the analysis of the TM RVs are shown in
   \fig{fig:GP_parameters}. In the RVs, the contribution from the GP reaches 2.3
   \ms, well above the amplitudes of either of the two planet signals. In the
   FWHM, the GP's contribution is 9.4 \ms.

   The rotation period of the star, parameterised by $\eta_3$, is constrained at
   a value of 85.1 days (posterior median), with a 1-$\sigma$ uncertainty below
   1 day. This value is close to half of the estimated timescale of evolution,
   $\eta_2 = 151$ days, suggesting that the active regions at the surface of the
   star (or at least the activity signal itself) evolve over a timescale of two
   rotation periods, similarly to what is observed in the Sun.
   The relatively large uncertainty on $\eta_2$ could be related to our
   assumption of sharing this parameter between RVs and FWHM, which might not be
   strictly valid (and indeed the same is true for $\eta_4$). We discuss briefly
   this assumption in Appendix \ref{app:covariances}.

   Finally, we note that the orbital period prior used in these analyses does
   not allow for the detection of \proxima c, which has an orbital period of 1900
   days \citep{Damasso2020a}. If the signal from this planet is present in the
   ESPRESSO dataset, we would expect it to be absorbed by the quadratic RV
   trend considered in our models. Nevertheless, for both the CCF and TM RVs,
   the coefficients of the quadratic trend are found to be compatible with zero.
   The limited time span, together with the inclusion of RV offsets between the
   three subsets of data makes the detection of \proxima c using only the
   currently available ESPRESSO data very challenging, in line with prediction
   from \citet{Damasso2020}. A complete analysis of the ESPRESSO data
   together with other RV datasets could shed light on the presence of this
   planet, but may require a more general activity model that includes long-term
   activity variations such as the magnetic cycle detected in \proxima
   \citep[see e.g.][]{SuarezMascareno2016,Wargelin2017}.

\section{Assessing the nature of the 5-day signal}
\label{sec:planet-nature}

   The RV analysis in the previous section included a GP component to model
   activity-induced RV variations and used the FWHM of the CCF as an activity
   indicator. Nevertheless, we cannot discard the possibility that the 5-day
   signal originates from an incomplete model of either stellar activity or
   instrumental systematics, or both. 
   In this section we study the evolution of the 5-day signal as the number of
   ESPRESSO RVs increased, look at other activity indicators to confirm the
   planetary nature of the signal and study the chromatic behaviour of the RVs.

   \subsection{Evolution with number of points}

      A planetary signal gives rise to strictly periodic RV variations and its
      stationary nature implies that it should become more significant as the
      number of observations increases. Using the same model and priors as in
      Sect. \ref{sec:rv-analysis}, we analysed the datasets built from the
      first $n$ ESPRESSO observations, with $n=30, 40 \ldots, 100, 114$. We note
      that this means the upper limit of the prior for the orbital periods
      increases with $n$, but this is of little consequence for this analysis
      given that the periodicities of interest are well covered already on the
      shortest dataset.

\begin{figure}
         \centering
         \includegraphics[width=\colfigsize]{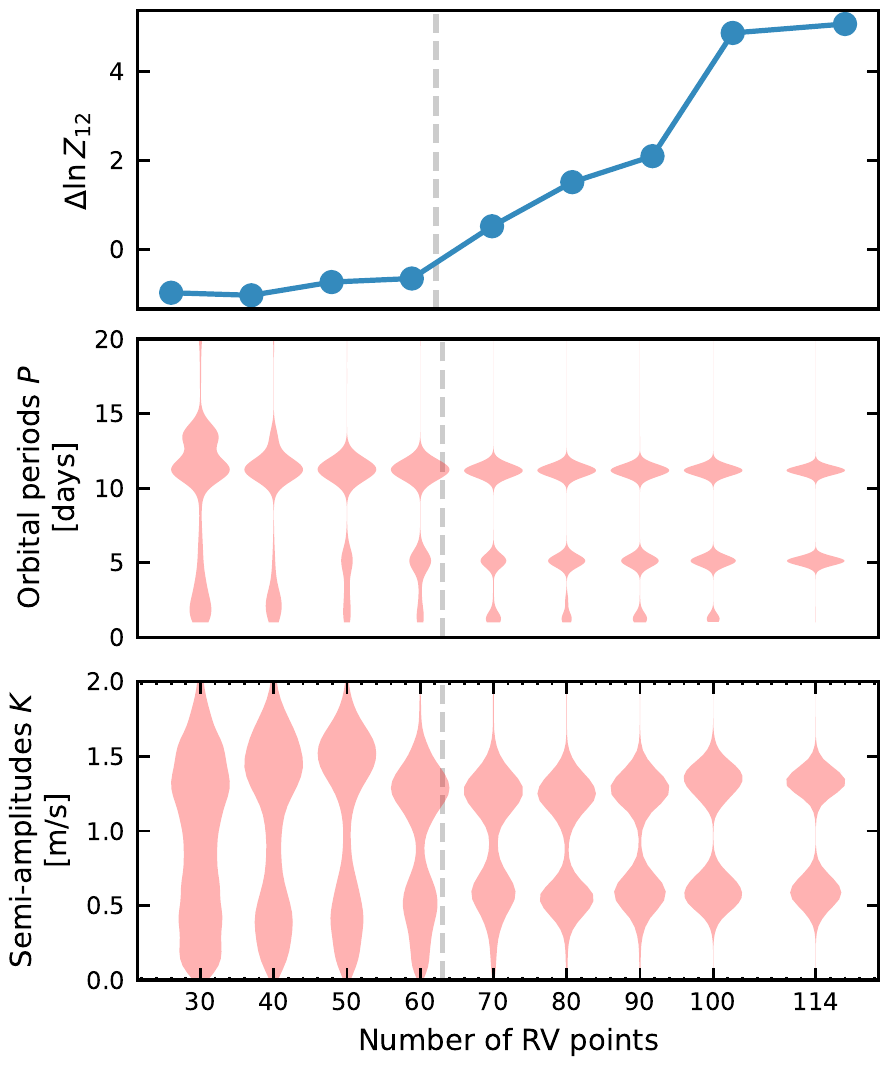}
         \caption{Evolution of the Bayes factor between models with one and two
         Keplerians (top) and of the posterior distribution for the orbital
         periods (middle) and semi-amplitudes (bottom) with an increasing number
         of ESPRESSO observations. The dashed vertical line corresponds to the
         number of observations available to \ct{SuarezMascareno2020}.}
         \label{fig:bf_K_P_violins}
      \end{figure}

      The results are shown in \fig{fig:bf_K_P_violins}, which displays the
      evolution of the Bayes factor between the models with $N_p=1$ and $N_p=2$
      together with the posterior distributions for the orbital periods and
      semi-amplitudes in the two-planet model, for increasing values of $n$. The
      Bayes factor increases monotonically until it reaches the value obtained
      for the full dataset. We note the large increase after the initial
      observing campaign presented in \ct{SuarezMascareno2020} and, after about
      90 observations, when we adapted the observing strategy to mitigate
      aliasing by obtaining two RV points per night when possible.
      
      It is clear that the posterior probability of the 5-day signal increases
      with the number of RVs in the dataset. The 11-day signal shows the same
      behaviour, even though its posterior is already well defined with fewer
      data points, due to the larger amplitude. At the same time, the
      semi-amplitudes of both signals also become better constrained as $n$
      increases, as seen in the bottom panel of \fig{fig:bf_K_P_violins}. These
      results provide evidence that the 5-day signal is stable over time, as
      expected for a planetary origin.

   \subsection{Periodicities in activity indicators}

   \begin{figure}
      \centering
      \includegraphics[width=\colfigsize]{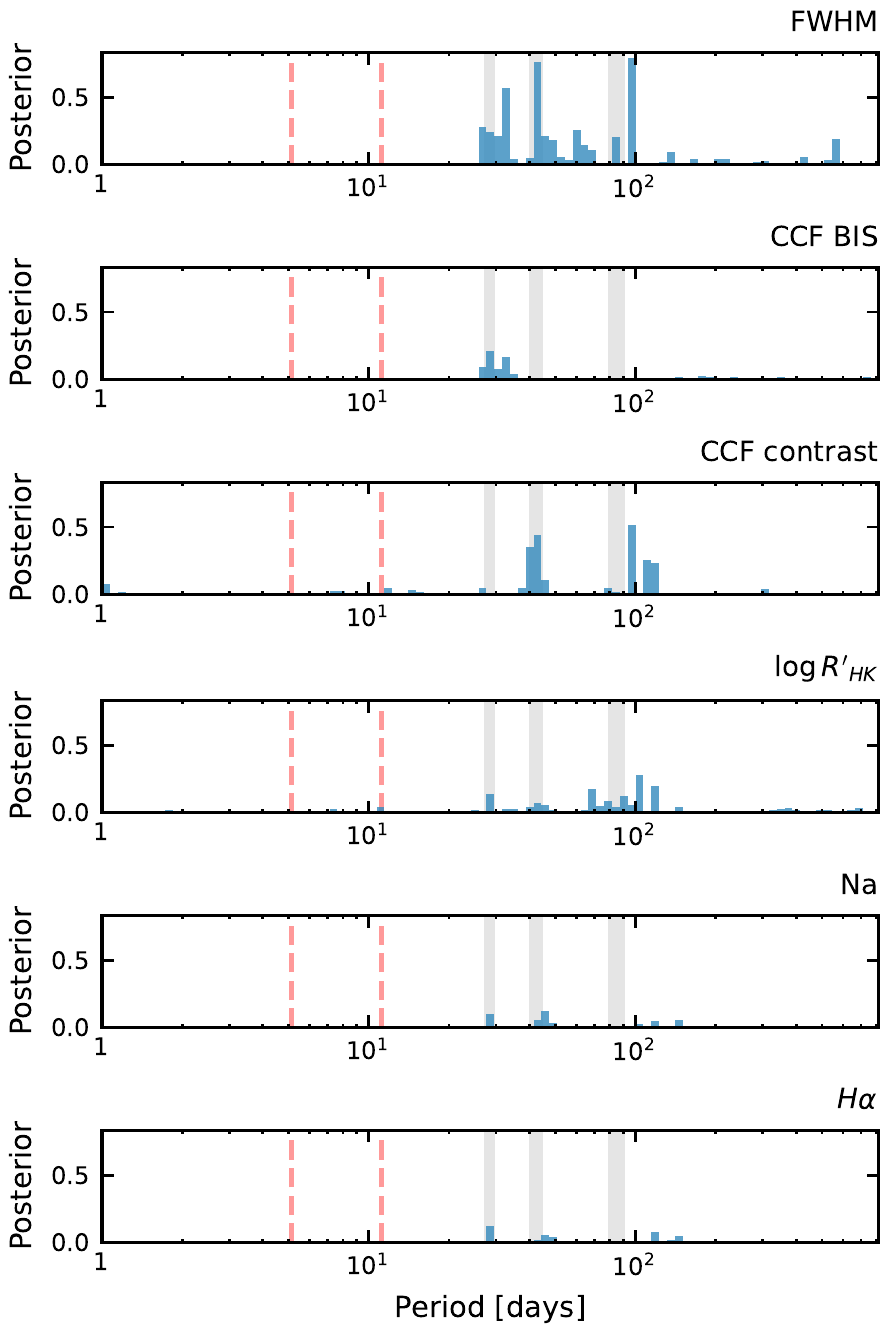}
      \caption{Posterior distributions for the periods of sinusoids in the
      analysis of activity indicators. The distributions are normalised and
      shown on the same vertical scale, which allows for a visual comparison of
      the significance of the periodicities. For example, the CCF contrast shows
      more significant periodicities (close to 100 and 40 days) than the
      H$\alpha$ index. The stellar rotation period and its first and second
      harmonics are denoted with filled lines, and the 11-day and 5-day signals
      detected in the RVs with dashed red lines.}
      \label{fig:periodograms}
   \end{figure}
 
   Besides the FWHM, a number of other activity indicators are available from
   the CCF or from individual spectral lines and can be used to identify
   activity signals. Moreover, all the indicators are obtained simultaneously to
   the RVs, so that the imprint of the observational sampling, if present,
   should appear similarly in each time series.

   Since any activity-related signal that might be present in the indicators is
   likely not periodic, the typical Lomb-Scargle periodogram might not be the
   ideal statistic to measure them. An arguably better model could use
   a QP GP to try to identify the stellar rotation signal and
   possibly an additional periodic component. But even this model might not be
   adequate, especially if it assumes the same priors as for the RVs, since the
   indicators can be sensitive to activity in different ways than the RVs, or be
   affected by instrumental effects.
   
   In addition, \proxima is known for its frequent flares
   \citep[e.g.][]{Davenport2016}, which introduces relatively strong outliers in
   the measurements of some activity-sensitive spectral lines, such as the Ca
   H\&K, H$\alpha$ or Na lines. These are indeed true outliers, in the sense
   that the estimated uncertainty of the affected points is not necessarily
   higher than that of unaffected observations.

   For these reasons, we decided to use a simple surrogate model to study the
   periodicities present in the activity indicators. We consider up to five
   sinusoidal signals, with periods constrained between 1 day and the time span
   of the observations. A Student's \emph{t} likelihood is used instead of the
   typical Gaussian in order to make the analysis more robust to outliers (the
   degrees of freedom of the distribution is a free parameter). We then plot the
   normalised posterior distribution for the five periods, which conveys
   information about the significance of each periodicity in the different time
   series. In practice, this results in a `cleaned' and robust version of the
   periodogram that retains only the most significant periods.

   The results are shown in \fig{fig:periodograms} for the FWHM, CCF bisector,
   CCF contrast, \logrhk, Na index, and H$\alpha$ index. Notably, the FWHM,
   \logrhk, and CCF contrast show significant periodicities close to the stellar
   rotation period and its first harmonic and all indicators show more or less
   significant periodicities at the second harmonic, close to 28 days. No clear
   peaks are seen around 11 or 5 days, suggesting that the two signals detected
   in the RVs are not related to activity, at least to the extent it is captured
   by the indicators. Again, these results add to the conclusion that the two
   signals are better explained as due to planetary companions.

   \subsection{Chromatic RV variations}

   As shown in \ct{SuarezMascareno2020}, the wide wavelength coverage of
   ESPRESSO, together with the collecting power of the VLT, allow for the
   derivation of so-called chromatic RVs by dividing the spectra into a few
   wavelength bins. \ct{SuarezMascareno2020} divided the ESPRESSO spectra into
   blue (440-570 nm), green (570-690 nm), and red (730-790 nm) bins, chosen to
   guarantee a similar RV precision. We select the same spectral orders to
   calculate chromatic RVs within the TM approach.

   The advantage of calculating chromatic RVs with the TM approach comes from
   the improved RV precision obtained when using this technique (see
   \fig{fig:plot_DRS_TM}). The blue, green, and red chromatic TM RVs show median
   RV uncertainties of 29, 24, and 22 \cms, respectively. In contrast, using the
   CCFs (and the same spectral orders) we achieve median uncertainties of 47,
   48, and 48 \cms for the three wavelength regions. The latter are comparable
   to the semi-amplitude measured for the 5-day signal, making a chromatic
   analysis of this signal challenging with the CCF RVs but possible with the
   TM RVs.

   \begin{figure*}
      \centering
      \includegraphics[width=0.95\hsize]{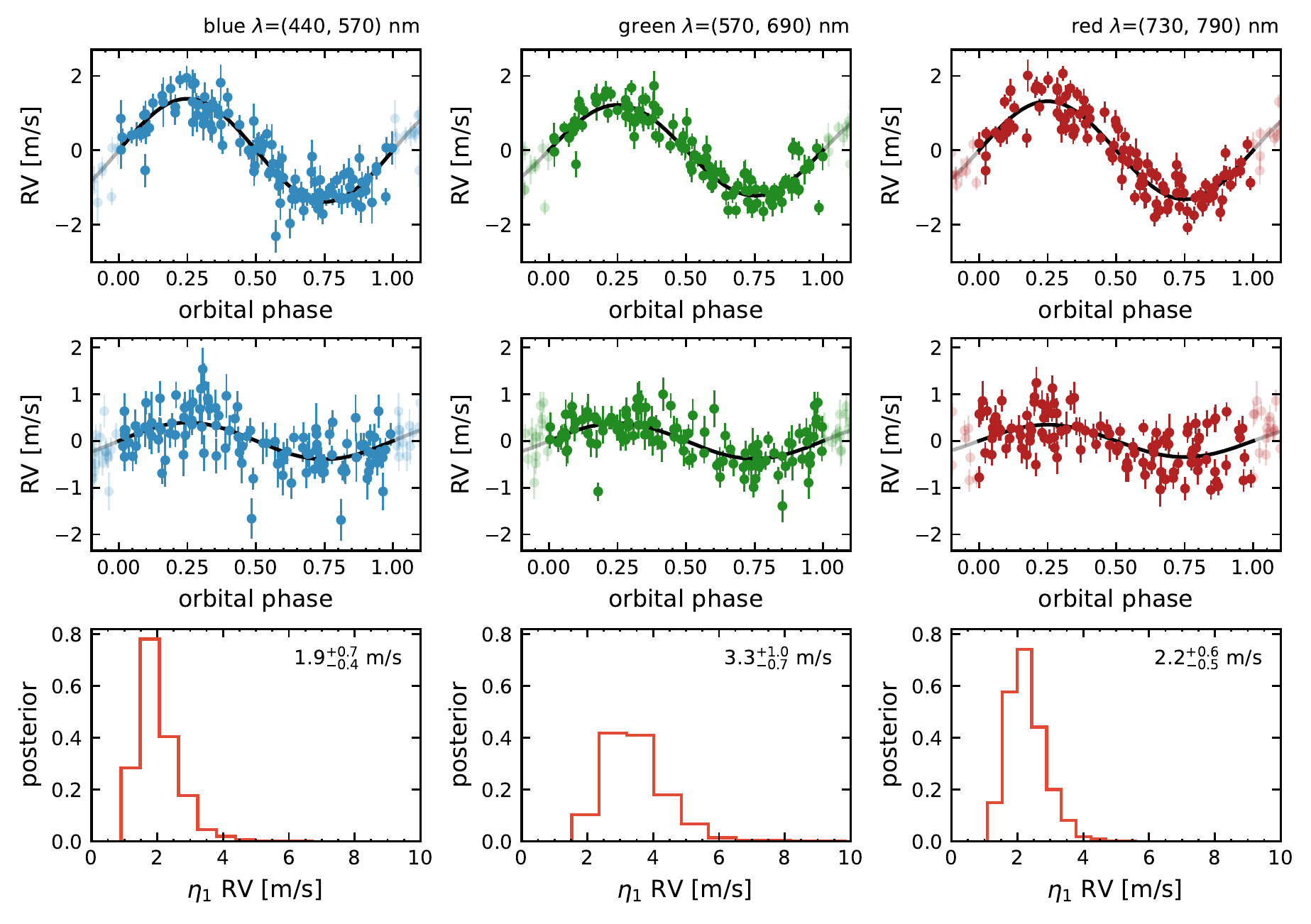}
      \caption{Analysis of chromatic TM RVs. The three columns show results for
      the blue, green, and red regions, corresponding to the wavelengths between
      440-570 nm, 570-690 nm, and 730-790 nm, respectively. In the top and
      middle panels, we show the ESPRESSO RVs phase-folded at the 11-day and
      5-day periods, together with the maximum likelihood solution for each
      signal. The bottom panels show the posteriors for the GP amplitude,
      $\eta_1$ RV, with the posterior median and 68\% quantiles as labels.}
      \label{fig:chromatic}
   \end{figure*}

   We analysed the blue, green, and red TM RV datasets individually, with a
   similar model to that of Sect. \ref{sec:rv-analysis}. The only difference
   is that here we assume only circular orbits, to minimise the number of
   parameters in the model and thus decrease the sensitivity to the larger
   contribution from photon noise to the error budget. The results are shown in
   \fig{fig:chromatic}. The top and middle panels show the phase curves of the
   11-day and 5-day signals, respectively, while the bottom panels show the
   posterior distributions for $\eta_1$ RV and $\eta_1$ FWHM.

   In all three wavelength bins the estimated semi-amplitudes are similar,
   varying by 16 \cms and 5 \cms for the 11-day and 5-day signals, respectively
   (about 13\% in both cases). A variation with wavelength would have been
   expected from a signal that was caused by a stellar spot
   \citep[e.g.][]{Desort2007,Figueira2010}. Nevertheless, the posterior
   probability for 5-day signal is lower in the red RVs, resulting in a less
   significant detection in this wavelength bin. This could be explained by a
   larger telluric contamination or a more complicated activity signal.
   
   Indeed we find hints that the amplitude of the activity signal in the RVs is
   larger in the green and red bins than in the blue, at least as captured by
   the GP model (see bottom panels in \fig{fig:chromatic}). However, the
   posterior estimates are mostly compatible, so this result should not be
   over-interpreted. In summary, this analysis of chromatic RVs provides us with
   additional confidence that the 5-day signal is of planetary origin.

\section{Discussion and conclusions}
\label{sec:discussion}

   In this work we have analysed a set of 114 ESPRESSO observations of \proxima.
   The imprint of \proxima b is the clearest signal seen in the ESPRESSO RVs (as
   in \ct{SuarezMascareno2020}) and is detected with very high significance in
   both the CCF and TM datasets. The orbital parameters are consistent in the
   two analyses, even if the TM RVs suggest a slightly smaller semi-amplitude.
   This smaller amplitude is closer to the one found by \ct{SuarezMascareno2020}
   from the combined analysis of ESPRESSO, HARPS, and UVES data. Using the TM
   RVs, the minimum mass of this planet is estimated as $1.07\pm0.06$\Mearth, with
   a relative uncertainty below 6\% (see Table \ref{tab:post}) when taking the
   uncertainty in the stellar mass into account.

   A significant 5.12-day signal is detected in the ESPRESSO data, which we
   attribute to a planet candidate, \proxima d, first found in
   \ct{SuarezMascareno2020}. We detect the signal with a slightly higher
   significance on the CCF RVs than on the TM RVs (Table \ref{tab:evidences}). 
   It is unlikely that this signal is directly connected to the stellar
   variability at the rotation period, which we estimate to be 84.5 days. First,
   the Keplerian signal becomes more significant as ESPRESSO observations are
   added, with its amplitude and period becoming increasingly better
   constrained, in a similar way to those of the 11-day signal. Second, none of
   the activity indicators we analysed show periodicities at either 5 or 11
   days, though they do vary with the rotation period and its harmonics.
   Finally, the 5-day signal shows a wavelength-independent amplitude within the
   range observed by ESPRESSO. These tests show evidence for the stationary and
   achromatic nature of the signal and are thus strong arguments in favour of
   the planetary nature of \proxima d.

   The estimated amplitude and eccentricity of \proxima d are smaller on the TM
   RVs and lead to an estimate for the minimum mass of $0.26 \pm 0.05$~\Mearth.
   This planet orbits closer to the star (at 0.02885 au) than to the inner edge
   of the HZ (see Table \ref{tab:parameters}), and it is the lightest planet
   detected with the RV method so far. 

   Together with other recent detections \citep{Demangeon2021,Lillo-Box2021},
   our results showcase the extreme RV precision systematically achieved with
   ESPRESSO. Even in the presence of stellar activity signals causing RV
   variations of the order of \ms, it is now possible to detect and measure
   precise masses for very low-mass planets that induce RV signals of only a few
   tens of \cms. 

   Our results bring new interest to the planetary system around \proxima. The
   star is now the likely host of three low-mass planets. The habitability
   conditions of \proxima b, which orbits within the HZ of the star, have been
   extensively studied
   \citep[e.g.][]{Barnes2017,Ribas2016,Turbet2016,Meadows2018}. On the other
   hand, the candidate \proxima d orbits much closer to the star and outside the
   HZ range. 

   The atmospheric properties of these planets may be affected by stellar
   magnetic activity \citep[e.g.][]{Garraffo2016,Garcia-Sage2017}. Recently,
   \citet{Klein2021} studied the extended magnetosphere of \proxima using
   circular polarisation spectra measured with HARPS-Pol and estimated the
   typical radius of the spherical Alfvén surface to be about 25 $R_*$. They
   concluded that \proxima b, which orbits at a distance of more than $70\:R_*$,
   should lie in the super-Alfvénic regime, with no direct star–planet magnetic
   connection. This conclusion was also supported by \citet{Kavanagh2021}. The
   inner planet, planet d, orbits at about $40\:R_*$, still in the
   super-Alfvénic regime.

   With the small stellar radius, \proxima d has a transit probability just
   above 2\%. Its equilibrium temperature may reach 360K, assuming a Bond albedo
   of 0.3 \citep[e.g.][]{Seager2010}. From the planetary properties and stellar
   parameters, we can estimate a planetary radius of $0.81 \pm 0.08$ \Rearth using
   the random forest model from \citet{Ulmer-Moll2019}, leading to a transit
   depth of about 0.3\% (approximately half that of \proxima b;
   \citealt{Anglada-Escude2016}). A transit detection would allow a precise
   measurement of the planetary radius and could place constraints on the planet
   bulk density and possible atmosphere. However, a transit is unlikely given
   that \proxima b has not been found to transit \citep[see][and references
   therein]{Jenkins2019a} and that transit events at periods below 5 days and
   depths above 3 mmag have been ruled out \citep{Feliz2019,Vida2019}.

   While the currently available ESPRESSO data do not allow for the detection of
   \proxima c, a full characterisation of the system may be possible with
   continued spectroscopic observations or with \emph{Gaia} astrometry
   \citep{Damasso2020,Damasso2020a}. A complete analysis of the available UVES,
   HARPS, and ESPRESSO datasets \citep[see][and references
   therein]{Damasso2020a,Anglada-Escude2016}, perhaps after a homogeneous RV
   derivation with TM, would further constrain the orbits of all the planets in
   the system. Given that this combined dataset would span several years, a more
   comprehensive stellar activity model (maybe including a long-term magnetic
   cycle) would have to be used, still extracting information from appropriate
   activity indicators. In summary, further observations of \proxima will help
   to complete our understanding of this multi-planetary system and may even
   unveil the presence of additional planets.

\begin{acknowledgements}
   The authors acknowledge the ESPRESSO project team for its effort and
   dedication in building the ESPRESSO instrument.

   This work was supported by FCT - Fundação para a Ciência e Tecnologia -
   through national funds and by FEDER through COMPETE2020 - Programa
   Operacional Competitividade e Internacionalização by these grants:
   UID/FIS/04434/2019; UIDB/04434/2020; UIDP/04434/2020;
   PTDC/FIS-AST/32113/2017 and POCI-01-0145-FEDER-032113;
   PTDC/FIS-AST/28953/2017 and POCI-01-0145-FEDER-028953;
   PTDC/FIS-AST/28987/2017 and POCI-01-0145-FEDER-028987;
   PTDC/FIS-AST/30389/2017 and POCI-01-0145-FEDER-030389.

   J.\,P.\,F. and O.\,D. are supported in the form of work contracts funded by
   national funds through FCT with the references DL 57/2016/CP1364/CT0005
   and DL 57/2016/CP1364/CT0004.

   A.\,M.\,S.  acknowledges  support  from FCT through  the  Fellowship
   2020.05387.BD and POCH/FSE(EC).

   A.\,S.\,M. acknowledges financial support from the Spanish Ministry of
   Science and Innovation (MICINN) under 2018 Juan de la Cierva program
   IJC2018-035229-I. 
   A.\,S.\,M., J.\,I.\,G.\,H., and R.\,R. acknowledge financial support from
   the MICINN project PID2020-117493GB-I00 and from the Government of the
   Canary Islands project ProID2020010129. J.\,I.\,G.\,H. also acknowledges
   financial support from the Spanish MICINN under 2013 Ram\'on y Cajal
   program RYC-2013-14875.

   Y.\,A. acknowledges the support of the Swiss National Fund under grant
   200020\_172746.

   The INAF authors acknowledge financial support of the Italian Ministry of
   Education, University, and Research with PRIN 201278X4FL and the "Progetti
   Premiali" funding scheme.

   J.\,L-B. acknowledges financial support received from ``la Caixa''
   Foundation (ID 100010434) and from the European Union’s Horizon 2020
   research and innovation programme under the Marie Skłodowska-Curie grant
   agreement No 847648, with fellowship code LCF/BQ/PI20/11760023.

   This work has been carried out within the framework of the NCCR PlanetS
   supported by the Swiss National Science Foundation. This project has
   received funding from the European Research Council (ERC) under the
   following European Union’s Horizon 2020 research and innovation
   programmes: grant agreement SCORE No 851555 and project {\sc Four Aces};
   grant agreement No 724427.
   D.\,E. acknowledges financial support from the Swiss National Science
   Foundation for project 200021\_200726.

   V.\,A. acknowledges the support from FCT through Investigador FCT contract
   nr. IF/00650/2015/CP1273/CT0001.
   
   N.\,J.\,N. acknowledges support form the following projects: UIDB/04434/2020
   \& UIDP/04434/2020, CERN/FIS-PAR/0037/2019, PTDC/FIS-OUT/29048/2017,
   COMPETE2020: POCI-01-0145-FEDER-028987 \& FCT: PTDC/FIS-AST/28987/2017.

   R.\,A. is a Trottier Postdoctoral Fellow and acknowledges support from the
   Trottier Family Foundation. This work was supported in part through a
   grant from FRQNT.

   This research has made use of the SIMBAD database, the VizieR catalogue
   operated at CDS, France, and the DACE platform developed in the frame of
   PlanetS (https://dace.unige.ch).

\end{acknowledgements}

\bibliographystyle{aa}
\bibliography{full_abrev}

\begin{thebibliography}{86}
\expandafter\ifx\csname natexlab\endcsname\relax\def\natexlab#1{#1}\fi

\bibitem[{{Anglada-Escud{\'e}} {et~al.}(2016){Anglada-Escud{\'e}}, Amado,
  Barnes, Berdi{\~n}as, Butler, Coleman, {de la Cueva}, Dreizler, Endl,
  Giesers, Jeffers, Jenkins, Jones, Kiraga, K{\"u}rster,
  {L{\'o}pez-Gonz{\'a}lez}, Marvin, Morales, Morin, Nelson, Ortiz, Ofir,
  Paardekooper, Reiners, Rodr{\'i}guez, {Rodr{\'i}guez-L{\'o}pez}, Sarmiento,
  Strachan, Tsapras, Tuomi, \& Zechmeister}]{Anglada-Escude2016}
{Anglada-Escud{\'e}}, G., Amado, P.~J., Barnes, J., {et~al.} 2016, \nat, 536,
  437

\bibitem[{{Anglada-Escud{\'e}} \& Butler(2012)}]{Anglada-Escude2012}
{Anglada-Escud{\'e}}, G. \& Butler, R.~P. 2012, The Astrophysical Journal
  Supplement Series, 200, 15

\bibitem[{Artigau {et~al.}(2014)Artigau, Kouach, Donati, Doyon, Delfosse,
  Baratchart, Lacombe, Moutou, Rabou, Par{\`e}s, Micheau, Thibault, Reshetov,
  Dubois, Hernandez, Vall{\'e}e, Wang, Dolon, Pepe, Bouchy, Striebig,
  H{\'e}nault, Loop, Saddlemyer, Barrick, Vermeulen, Dupieux, H{\'e}brard,
  Boisse, Martioli, Alencar, {do Nascimento}, \& Figueira}]{Artigau2014}
Artigau, {\'E}., Kouach, D., Donati, J.-F., {et~al.} 2014, in Ground-Based and
  {{Airborne Instrumentation}} for {{Astronomy V}}, ed. S.~K. Ramsay, I.~S.
  McLean, \& H.~Takami, 914715

\bibitem[{Baglin {et~al.}(2006)Baglin, Auvergne, Barge, Deleuil, Catala,
  Michel, Weiss, \& {COROT Team}}]{Baglin2006}
Baglin, A., Auvergne, M., Barge, P., {et~al.} 2006, 1306, 33

\bibitem[{Baranne {et~al.}(1996)Baranne, Queloz, Mayor, Adrianzyk, Knispel,
  Kohler, Lacroix, Meunier, Rimbaud, \& Vin}]{Baranne1996}
Baranne, A., Queloz, D., Mayor, M., {et~al.} 1996, Astronomy and Astrophysics
  Supplement Series, 119, 373

\bibitem[{Barnes {et~al.}(2017)Barnes, Jeffers, {Anglada-Escud{\'e}}, Haswell,
  Jones, Tuomi, Feng, Jenkins, \& Petit}]{Barnes2017}
Barnes, J.~R., Jeffers, S.~V., {Anglada-Escud{\'e}}, G., {et~al.} 2017, \mnras,
  466, 1733

\bibitem[{Benedict \& McArthur(2020)}]{Benedict2020}
Benedict, G.~F. \& McArthur, B.~E. 2020, Research Notes of the American
  Astronomical Society, 4, 86

\bibitem[{Bertaux {et~al.}(2014)Bertaux, Lallement, Ferron, Boonne, \&
  Bodichon}]{Bertaux2014}
Bertaux, J.~L., Lallement, R., Ferron, S., Boonne, C., \& Bodichon, R. 2014,
  \aap, 564, A46

\bibitem[{Bessell(1991)}]{Bessell1991}
Bessell, M.~S. 1991, \aj, 101, 662

\bibitem[{Bleistein \& Handelsman(1986)}]{Bleistein1986}
Bleistein, N. \& Handelsman, R.~A. 1986, Asymptotic Expansions of Integrals
  ({New York}: {Dover Publications})

\bibitem[{Bonfils {et~al.}(2018)Bonfils, {Astudillo-Defru}, D{\'i}az, Almenara,
  Forveille, Bouchy, Delfosse, Lovis, Mayor, Murgas, Pepe, Santos,
  S{\'e}gransan, Udry, \& W{\"u}nsche}]{Bonfils2018}
Bonfils, X., {Astudillo-Defru}, N., D{\'i}az, R., {et~al.} 2018, \aap, 613, A25

\bibitem[{Borucki(2016)}]{Borucki2016}
Borucki, W.~J. 2016, Reports on Progress in Physics, 79, 036901

\bibitem[{Bouchy {et~al.}(2001)Bouchy, Pepe, \& Queloz}]{Bouchy2001}
Bouchy, F., Pepe, F., \& Queloz, D. 2001, \aap, 374, 733

\bibitem[{Boyajian {et~al.}(2012)Boyajian, {von Braun}, {van Belle}, McAlister,
  {ten Brummelaar}, Kane, Muirhead, Jones, White, Schaefer, Ciardi, Henry,
  {L{\'o}pez-Morales}, Ridgway, Gies, Jao, {Rojas-Ayala}, Parks, Sturmann,
  Sturmann, Turner, Farrington, Goldfinger, \& Berger}]{Boyajian2012}
Boyajian, T.~S., {von Braun}, K., {van Belle}, G., {et~al.} 2012, \apj, 757,
  112

\bibitem[{Brewer(2014)}]{Brewer2014}
Brewer, B.~J. 2014 [\eprint[arXiv]{1411.3921}]

\bibitem[{Brewer {et~al.}(2011)Brewer, P{\'a}rtay, \& Cs{\'a}nyi}]{Brewer2011}
Brewer, B.~J., P{\'a}rtay, L.~B., \& Cs{\'a}nyi, G. 2011, Statistics and
  Computing, 21, 649

\bibitem[{Broeg {et~al.}(2013)Broeg, Fortier, Ehrenreich, Alibert, Baumjohann,
  Benz, Deleuil, Gillon, Ivanov, Liseau, Meyer, Oloffson, Pagano, Piotto,
  Pollacco, Queloz, Ragazzoni, Renotte, Steller, Thomas, \& {the CHEOPS
  team}}]{Broeg2013}
Broeg, C., Fortier, A., Ehrenreich, D., {et~al.} 2013, EPJ Web of Conferences,
  47, 03005

\bibitem[{Collier~Cameron {et~al.}(2019)Collier~Cameron, Mortier, Phillips,
  Dumusque, Haywood, Langellier, Watson, Cegla, Costes, Charbonneau, Coffinet,
  Latham, {Lopez-Morales}, Malavolta, Maldonado, Micela, Milbourne, Molinari,
  Saar, Thompson, Buchschacher, Cecconi, Cosentino, Ghedina, Glenday, Gonzalez,
  Li, Lodi, Lovis, Pepe, Poretti, Rice, Sasselov, Sozzetti, Szentgyorgyi, Udry,
  \& Walsworth}]{CollierCameron2019}
Collier~Cameron, A., Mortier, A., Phillips, D., {et~al.} 2019, \mnras, 487,
  1082

\bibitem[{Cosentino {et~al.}(2012)Cosentino, Lovis, Pepe, Collier~Cameron,
  Latham, Molinari, Udry, Bezawada, Black, Born, Buchschacher, Charbonneau,
  Figueira, Fleury, Galli, Gallie, Gao, Ghedina, Gonzalez, Gonzalez, Guerra,
  Henry, Horne, Hughes, Kelly, Lodi, Lunney, Maire, Mayor, Micela, Ordway,
  Peacock, Phillips, Piotto, Pollacco, Queloz, Rice, Riverol, Riverol,
  San~Juan, Sasselov, Segransan, Sozzetti, Sosnowska, Stobie, Szentgyorgyi,
  Vick, \& Weber}]{Cosentino2012a}
Cosentino, R., Lovis, C., Pepe, F., {et~al.} 2012, 8446, 84461V

\bibitem[{Damasso \& Del~Sordo(2020)}]{Damasso2020}
Damasso, M. \& Del~Sordo, F. 2020, 494, 1387

\bibitem[{Damasso {et~al.}(2020)Damasso, Del~Sordo, {Anglada-Escud{\'e}},
  Giacobbe, Sozzetti, Morbidelli, Pojmanski, Barbato, Butler, Jones, Hambsch,
  Jenkins, {L{\'o}pez-Gonz{\'a}lez}, Morales, Pe{\~n}a~Rojas,
  {Rodr{\'i}guez-L{\'o}pez}, Rodr{\'i}guez, Amado, Anglada, Feng, \&
  G{\'o}mez}]{Damasso2020a}
Damasso, M., Del~Sordo, F., {Anglada-Escud{\'e}}, G., {et~al.} 2020, Science
  Advances, 6, eaax7467

\bibitem[{Davenport {et~al.}(2016)Davenport, Kipping, Sasselov, Matthews, \&
  Cameron}]{Davenport2016}
Davenport, J. R.~A., Kipping, D.~M., Sasselov, D., Matthews, J.~M., \& Cameron,
  C. 2016, \apj, 829, L31

\bibitem[{Delfosse {et~al.}(2000)Delfosse, Forveille, S{\'e}gransan, Beuzit,
  Udry, Perrier, \& Mayor}]{Delfosse2000}
Delfosse, X., Forveille, T., S{\'e}gransan, D., {et~al.} 2000, \aap, 364, 217

\bibitem[{Demangeon {et~al.}(2021)Demangeon, Osorio, Alibert, Barros,
  Adibekyan, Tabernero, {Antoniadis-Karnavas}, Camacho, Mascare{\~n}o, Oshagh,
  Micela, Sousa, Lovis, Pepe, Rebolo, Cristiani, Santos, Allart, Prieto,
  Bossini, Bouchy, Cabral, Damasso, Marcantonio, D'Odorico, Ehrenreich, Faria,
  Figueira, Santos, Haldemann, Hara, Hern{\'a}ndez, Lavie, {Lillo-Box}, Curto,
  Martins, M{\'e}gevand, Mehner, Molaro, Nunes, Pall{\'e}, Pasquini, Poretti,
  Sozzetti, \& Udry}]{Demangeon2021}
Demangeon, O. D.~S., Osorio, M. R.~Z., Alibert, Y., {et~al.} 2021, \aap, 653,
  A41

\bibitem[{Desort {et~al.}(2007)Desort, Lagrange, Galland, Udry, \&
  Mayor}]{Desort2007}
Desort, M., Lagrange, A.-M., Galland, F., Udry, S., \& Mayor, M. 2007, \aap,
  473, 983

\bibitem[{D{\'i}az {et~al.}(2019)D{\'i}az, Delfosse, Hobson, Boisse,
  {Astudillo-Defru}, Bonfils, Henry, Arnold, Bouchy, Bourrier, Brugger, Dalal,
  Deleuil, Demangeon, Dolon, Dumusque, Forveille, Hara, H{\'e}brard, Kiefer,
  Lopez, Mignon, Moreau, Mousis, Moutou, Pepe, Perruchot, Richaud, Santerne,
  Santos, Sottile, Stalport, S{\'e}gransan, Udry, Unger, \& Wilson}]{Diaz2019}
D{\'i}az, R.~F., Delfosse, X., Hobson, M.~J., {et~al.} 2019, \aap, 625, A17

\bibitem[{Ehrenreich {et~al.}(2020)Ehrenreich, Lovis, Allart, Zapatero~Osorio,
  Pepe, Cristiani, Rebolo, Santos, Borsa, Demangeon, Dumusque,
  Gonz{\'a}lez~Hern{\'a}ndez, {Casasayas-Barris}, S{\'e}gransan, Sousa, Abreu,
  Adibekyan, Affolter, Allende~Prieto, Alibert, Aliverti, Alves, Amate, Avila,
  Baldini, Bandy, Benz, Bianco, Bolmont, Bouchy, Bourrier, Broeg, Cabral,
  Calderone, Pall{\'e}, Cegla, Cirami, Coelho, Conconi, Coretti, Cumani,
  Cupani, Dekker, Delabre, Deiries, D'Odorico, Di~Marcantonio, Figueira,
  Fragoso, Genolet, Genoni, G{\'e}nova~Santos, Hara, Hughes, Iwert, Kerber,
  Knudstrup, Landoni, Lavie, Lizon, Lendl, Lo~Curto, Maire, Manescau, Martins,
  M{\'e}gevand, Mehner, Micela, Modigliani, Molaro, Monteiro, Monteiro,
  Moschetti, M{\"u}ller, Nunes, Oggioni, Oliveira, Pariani, Pasquini, Poretti,
  Rasilla, Redaelli, Riva, Santana~Tschudi, Santin, Santos, Segovia~Milla,
  Seidel, Sosnowska, Sozzetti, Span{\`o}, Su{\'a}rez~Mascare{\~n}o, Tabernero,
  Tenegi, Udry, Zanutta, \& Zerbi}]{Ehrenreich2020}
Ehrenreich, D., Lovis, C., Allart, R., {et~al.} 2020, \nat, 580, 597

\bibitem[{Faria {et~al.}(2018)Faria, Santos, Figueira, \& Brewer}]{Faria2018}
Faria, J.~P., Santos, N.~C., Figueira, P., \& Brewer, B.~J. 2018, JOSS, 3, 487

\bibitem[{Feliz {et~al.}(2019)Feliz, Blank, Collins, White, Stassun, Curtis,
  Hart, Kielkopf, Nelson, Relles, Stockdale, Jayawardene, Shankland, Reichart,
  Haislip, \& Kouprianov}]{Feliz2019}
Feliz, D.~L., Blank, D.~L., Collins, K.~A., {et~al.} 2019, \aj, 157, 226

\bibitem[{Feroz {et~al.}(2011)Feroz, Balan, \& Hobson}]{Feroz2011}
Feroz, F., Balan, S.~T., \& Hobson, M.~P. 2011, \mnras, 415, 3462

\bibitem[{Figueira {et~al.}(2010)Figueira, Marmier, Bonfils, {di Folco}, Udry,
  Santos, Lovis, M{\'e}gevand, Melo, Pepe, Queloz, S{\'e}gransan, Triaud, \&
  Viana~Almeida}]{Figueira2010}
Figueira, P., Marmier, M., Bonfils, X., {et~al.} 2010, \aap, 513, L8

\bibitem[{Fischer {et~al.}(2016)Fischer, {Anglada-Escude}, Arriagada, Baluev,
  Bean, Bouchy, Buchhave, Carroll, Chakraborty, Crepp, Dawson, Diddams,
  Dumusque, Eastman, Endl, Figueira, Ford, {Daniel Foreman-Mackey}, Fournier,
  F{\H u}r{\'e}sz, Gaudi, Gregory, Grundahl, Hatzes, H{\'e}brard, Herrero,
  Hogg, Howard, Johnson, {Paul Jorden}, Jurgenson, Latham, Laughlin, Loredo,
  Lovis, {Suvrath Mahadevan}, McCracken, Pepe, Perez, Phillips, Plavchan, {Lisa
  Prato}, Quirrenbach, Reiners, Robertson, Santos, Sawyer, {Damien Segransan},
  Sozzetti, Steinmetz, Szentgyorgyi, Udry, Valenti, Wang, Wittenmyer, \&
  Wright}]{Fischer2016}
Fischer, D.~A., {Anglada-Escude}, G., Arriagada, P., {et~al.} 2016, \pasp, 128,
  066001

\bibitem[{{Gaia Collaboration} {et~al.}(2016){Gaia Collaboration}, Brown,
  Vallenari, Prusti, {de Bruijne}, Mignard, Drimmel, Babusiaux, {Bailer-Jones},
  Bastian, Biermann, Evans, Eyer, Jansen, Jordi, Katz, Klioner, Lammers,
  Lindegren, Luri, O'Mullane, Panem, Pourbaix, Randich, Sartoretti, Siddiqui,
  Soubiran, Valette, {van Leeuwen}, Walton, Aerts, Arenou, Cropper, H{\o}g,
  Lattanzi, Grebel, Holland, Huc, Passot, Perryman, Bramante, Cacciari,
  Casta{\~n}eda, Chaoul, Cheek, De~Angeli, Fabricius, Guerra, Hern{\'a}ndez,
  {Jean-Antoine-Piccolo}, Masana, Messineo, Mowlavi, Nienartowicz,
  {Ord{\'o}{\~n}ez-Blanco}, Panuzzo, Portell, Richards, Riello, Seabroke,
  Tanga, Th{\'e}venin, Torra, Els, {Gracia-Abril}, Comoretto,
  {Garcia-Reinaldos}, Lock, Mercier, Altmann, Andrae, Astraatmadja,
  {Bellas-Velidis}, Benson, Berthier, Blomme, Busso, Carry, Cellino,
  Clementini, Cowell, Creevey, Cuypers, Davidson, De~Ridder, {de Torres},
  Delchambre, Dell'Oro, Ducourant, Fr{\'e}mat, {Garc{\'i}a-Torres}, Gosset,
  Halbwachs, Hambly, Harrison, Hauser, Hestroffer, Hodgkin, Huckle, Hutton,
  Jasniewicz, Jordan, Kontizas, Korn, Lanzafame, Manteiga, Moitinho, Muinonen,
  Osinde, Pancino, Pauwels, Petit, {Recio-Blanco}, Robin, Sarro, Siopis, Smith,
  Smith, Sozzetti, Thuillot, {van Reeven}, Viala, Abbas, Abreu~Aramburu,
  Accart, Aguado, Allan, Allasia, Altavilla, {\'A}lvarez, Alves, Anderson,
  Andrei, Anglada~Varela, Antiche, Antoja, Ant{\'o}n, Arcay, Bach, Baker,
  {Balaguer-N{\'u}{\~n}ez}, Barache, Barata, Barbier, Barblan, {Barrado y
  Navascu{\'e}s}, Barros, Barstow, Becciani, Bellazzini, Bello~Garc{\'i}a,
  Belokurov, Bendjoya, Berihuete, Bianchi, Bienaym{\'e}, Billebaud,
  Blagorodnova, {Blanco-Cuaresma}, Boch, Bombrun, Borrachero, Bouquillon,
  Bourda, Bouy, Bragaglia, Breddels, Brouillet, Br{\"u}semeister, Bucciarelli,
  Burgess, Burgon, Burlacu, Busonero, Buzzi, Caffau, Cambras, Campbell,
  Cancelliere, {Cantat-Gaudin}, Carlucci, Carrasco, Castellani, Charlot,
  Charnas, Chiavassa, Clotet, Cocozza, Collins, Costigan, Crifo, Cross, Crosta,
  Crowley, Dafonte, Damerdji, Dapergolas, David, David, De~Cat, {de Felice},
  {de Laverny}, De~Luise, De~March, {de Martino}, {de Souza}, Debosscher, {del
  Pozo}, Delbo, Delgado, Delgado, Di~Matteo, Diakite, Distefano, Dolding,
  Dos~Anjos, Drazinos, Duran, Dzigan, Edvardsson, Enke, Evans, Eynard~Bontemps,
  Fabre, Fabrizio, Faigler, Falc{\~a}o, Farr{\`a}s~Casas, Federici, Fedorets,
  {Fern{\'a}ndez-Hern{\'a}ndez}, Fernique, Fienga, Figueras, Filippi,
  Findeisen, Fonti, Fouesneau, Fraile, Fraser, Fuchs, Gai, Galleti, Galluccio,
  Garabato, {Garc{\'i}a-Sedano}, Garofalo, Garralda, Gavras, Gerssen, Geyer,
  Gilmore, Girona, Giuffrida, Gomes, {Gonz{\'a}lez-Marcos},
  {Gonz{\'a}lez-N{\'u}{\~n}ez}, {Gonz{\'a}lez-Vidal}, Granvik, Guerrier,
  Guillout, Guiraud, G{\'u}rpide, {Guti{\'e}rrez-S{\'a}nchez}, Guy, Haigron,
  Hatzidimitriou, Haywood, Heiter, Helmi, Hobbs, Hofmann, Holl, Holland, Hunt,
  Hypki, Icardi, Irwin, {Jevardat de Fombelle}, Jofr{\'e}, Jonker, Jorissen,
  Julbe, Karampelas, Kochoska, Kohley, Kolenberg, Kontizas, Koposov,
  Kordopatis, Koubsky, {Krone-Martins}, Kudryashova, Kull, Bachchan,
  {Lacoste-Seris}, Lanza, Lavigne, {Le Poncin-Lafitte}, Lebreton, Lebzelter,
  Leccia, Leclerc, {Lecoeur-Taibi}, Lemaitre, Lenhardt, Leroux, Liao, Licata,
  Lindstr{\o}m, Lister, Livanou, Lobel, L{\"o}ffler, L{\'o}pez, Lorenz,
  MacDonald, Magalh{\~a}es~Fernandes, Managau, Mann, Mantelet, Marchal,
  Marchant, Marconi, Marinoni, Marrese, Marschalk{\'o}, Marshall,
  {Mart{\'i}n-Fleitas}, Martino, Mary, Matijevi{\v c}, Mazeh, McMillan,
  Messina, Michalik, Millar, Miranda, Molina, Molinaro, Molinaro, Moln{\'a}r,
  Moniez, Montegriffo, Mor, Mora, Morbidelli, Morel, Morgenthaler, Morris,
  Mulone, Muraveva, Musella, Narbonne, Nelemans, Nicastro, Noval,
  Ord{\'e}novic, {Ordieres-Mer{\'e}}, Osborne, Pagani, Pagano, Pailler,
  Palacin, Palaversa, Parsons, Pecoraro, Pedrosa, Pentik{\"a}inen, Pichon,
  Piersimoni, Pineau, Plachy, Plum, Poujoulet, Pr{\v s}a, Pulone, Ragaini,
  Rago, Rambaux, {Ramos-Lerate}, Ranalli, Rauw, Read, Regibo, Reyl{\'e},
  Ribeiro, Rimoldini, Ripepi, Riva, Rixon, Roelens, {Romero-G{\'o}mez}, Rowell,
  Royer, {Ruiz-Dern}, Sadowski, Sagrist{\`a}~Sell{\'e}s, Sahlmann, Salgado,
  Salguero, Sarasso, Savietto, Schultheis, Sciacca, Segol, Segovia, Segransan,
  Shih, Smareglia, Smart, Solano, Solitro, Sordo, Soria~Nieto, Souchay, Spagna,
  Spoto, Stampa, Steele, Steidelm{\"u}ller, Stephenson, Stoev, Suess,
  S{\"u}veges, Surdej, Szabados, {Szegedi-Elek}, Tapiador, Taris, Tauran,
  Taylor, Teixeira, Terrett, Tingley, Trager, Turon, Ulla, Utrilla, Valentini,
  {van Elteren}, Van~Hemelryck, {van Leeuwen}, Varadi, Vecchiato, Veljanoski,
  Via, Vicente, Vogt, Voss, Votruba, Voutsinas, Walmsley, Weiler, Weingrill,
  Wevers, Wyrzykowski, Yoldas, {\v Z}erjal, Zucker, Zurbach, Zwitter, Alecu,
  Allen, Allende~Prieto, Amorim, {Anglada-Escud{\'e}}, Arsenijevic, Azaz, Balm,
  Beck, Bernstein, Bigot, Bijaoui, Blasco, Bonfigli, Bono, Boudreault, Bressan,
  Brown, Brunet, Bunclark, Buonanno, Butkevich, Carret, Carrion, Chemin,
  Ch{\'e}reau, Corcione, Darmigny, {de Boer}, {de Teodoro}, {de Zeeuw},
  Delle~Luche, Domingues, Dubath, Fodor, Fr{\'e}zouls, Fries, Fustes, Fyfe,
  Gallardo, Gallegos, Gardiol, Gebran, Gomboc, G{\'o}mez, Grux, Gueguen,
  Heyrovsky, Hoar, Iannicola, Isasi~Parache, Janotto, Joliet, Jonckheere, Keil,
  Kim, Klagyivik, Klar, Knude, Kochukhov, Kolka, Kos, Kutka, Lainey, LeBouquin,
  Liu, Loreggia, Makarov, Marseille, Martayan, {Martinez-Rubi}, Massart,
  Meynadier, Mignot, Munari, Nguyen, Nordlander, Ocvirk, O'Flaherty,
  Olias~Sanz, Ortiz, Osorio, Oszkiewicz, Ouzounis, Palmer, Park, Pasquato,
  Peltzer, Peralta, P{\'e}turaud, Pieniluoma, Pigozzi, Poels, Prat, Prod'homme,
  Raison, Rebordao, Risquez, {Rocca-Volmerange}, Rosen, {Ruiz-Fuertes}, Russo,
  Sembay, Serraller~Vizcaino, Short, Siebert, Silva, Sinachopoulos, Slezak,
  Soffel, Sosnowska, Strai{\v z}ys, {ter Linden}, Terrell, Theil, Tiede,
  Troisi, Tsalmantza, Tur, Vaccari, Vachier, Valles, Van~Hamme, Veltz,
  Virtanen, Wallut, Wichmann, Wilkinson, Ziaeepour, \&
  Zschocke}]{GaiaCollaboration2016}
{Gaia Collaboration}, Brown, A. G.~A., Vallenari, A., {et~al.} 2016, \aap, 595,
  A2

\bibitem[{{Garcia-Sage} {et~al.}(2017){Garcia-Sage}, Glocer, Drake, Gronoff, \&
  Cohen}]{Garcia-Sage2017}
{Garcia-Sage}, K., Glocer, A., Drake, J.~J., Gronoff, G., \& Cohen, O. 2017,
  \apj, 844, L13

\bibitem[{Garraffo {et~al.}(2016)Garraffo, Drake, \& Cohen}]{Garraffo2016}
Garraffo, C., Drake, J.~J., \& Cohen, O. 2016, \apj, 833, L4

\bibitem[{{Gomes da Silva} {et~al.}(2018){Gomes da Silva}, Figueira, Santos, \&
  Faria}]{GomesdaSilva2018}
{Gomes da Silva}, J., Figueira, P., Santos, N.~C., \& Faria, J.~P. 2018, JOSS,
  3, 667

\bibitem[{Gratton {et~al.}(2020)Gratton, Zurlo, Le~Coroller, Damasso,
  Del~Sordo, Langlois, Mesa, Milli, Chauvin, Desidera, Hagelberg, Lagadec,
  Vigan, Boccaletti, Bonnefoy, Brandner, Brown, Cantalloube, Delorme, D'Orazi,
  Feldt, Galicher, Henning, Janson, Kervella, Lagrange, Lazzoni, Ligi, Maire,
  M{\'e}nard, Meyer, Mugnier, Potier, Rickman, Rodet, Romero, Schmidt, Sissa,
  Sozzetti, Szul{\'a}gyi, Wahhaj, Antichi, Fusco, Stadler, Suarez, \&
  Wildi}]{Gratton2020}
Gratton, R., Zurlo, A., Le~Coroller, H., {et~al.} 2020, 638, A120

\bibitem[{Gregory(2005)}]{Gregory2005}
Gregory, P.~C. 2005, \apj, 631, 1198

\bibitem[{Grunblatt {et~al.}(2015)Grunblatt, Howard, \&
  Haywood}]{Grunblatt2015}
Grunblatt, S.~K., Howard, A.~W., \& Haywood, R.~D. 2015, \apj, 808, 127

\bibitem[{Hara {et~al.}(2021)Hara, Unger, Delisle, D{\'i}az, \&
  S{\'e}gransan}]{Hara2021}
Hara, N.~C., Unger, N., Delisle, J.-B., D{\'i}az, R., \& S{\'e}gransan, D.
  2021, arXiv:2105.06995 [astro-ph, stat] [\eprint[arXiv]{2105.06995}]

\bibitem[{Haywood {et~al.}(2014)Haywood, Collier~Cameron, Queloz, Barros,
  Deleuil, Fares, Gillon, Lanza, Lovis, Moutou, Pepe, Pollacco, Santerne,
  S{\'e}gransan, \& Unruh}]{Haywood2014}
Haywood, R.~D., Collier~Cameron, A., Queloz, D., {et~al.} 2014, \mnras, 443,
  2517

\bibitem[{Hojjatpanah {et~al.}(2019)Hojjatpanah, Figueira, Santos, Adibekyan,
  Sousa, {Delgado-Mena}, Alibert, Cristiani, Gonz{\'a}lez~Hern{\'a}ndez, Lanza,
  Di~Marcantonio, Martins, Micela, Molaro, Neves, Oshagh, Pepe, Poretti,
  {Rojas-Ayala}, Rebolo, Su{\'a}rez~Mascare{\~n}o, \&
  Zapatero~Osorio}]{Hojjatpanah2019}
Hojjatpanah, S., Figueira, P., Santos, N.~C., {et~al.} 2019, \aap, 629, A80

\bibitem[{Jao {et~al.}(2014)Jao, Henry, Subasavage, Winters, Gies, Riedel, \&
  Ianna}]{Jao2014}
Jao, W.-C., Henry, T.~J., Subasavage, J.~P., {et~al.} 2014, \aj, 147, 21

\bibitem[{Jeffers {et~al.}(2020)Jeffers, Dreizler, Barnes, Haswell, Nelson,
  Rodr{\'i}guez, {L{\'o}pez-Gonz{\'a}lez}, Morales, Luque, Zechmeister, Vogt,
  Jenkins, Palle, {\~N}as, Coleman, D{\'i}az, Ribas, Jones, Butler, Tinney,
  Bailey, Carter, O'Toole, Wittenmyer, Crane, Feng, Shectman, Teske, Reiners,
  Amado, \& {Anglada-Escud{\'e}}}]{Jeffers2020}
Jeffers, S.~V., Dreizler, S., Barnes, J.~R., {et~al.} 2020, Science, 368, 1477

\bibitem[{Jenkins {et~al.}(2019)Jenkins, Harrington, Challener, Kurtovic,
  Ramirez, Pe{\~n}a, McIntyre, Himes, Rodr{\'i}guez, {Anglada-Escud{\'e}},
  Dreizler, Ofir, Rojas, Ribas, Rojo, Kipping, Butler, Amado,
  {Rodr{\'i}guez-L{\'o}pez}, Kempton, Palle, \& Murgas}]{Jenkins2019a}
Jenkins, J.~S., Harrington, J., Challener, R.~C., {et~al.} 2019, \mnras, 487,
  268

\bibitem[{Kass \& Raftery(1995)}]{Kass1995}
Kass, R.~E. \& Raftery, A.~E. 1995, Journal of the American Statistical
  Association, 90, 773

\bibitem[{Kavanagh {et~al.}(2021)Kavanagh, Vidotto, Klein, Jardine, Donati, \&
  {\'O}~Fionnag{\'a}in}]{Kavanagh2021}
Kavanagh, R.~D., Vidotto, A.~A., Klein, B., {et~al.} 2021, 504, 1511

\bibitem[{Kervella {et~al.}(2020)Kervella, Arenou, \& Schneider}]{Kervella2020}
Kervella, P., Arenou, F., \& Schneider, J. 2020, \aap, 635, L14

\bibitem[{Kipping(2013)}]{Kipping2013}
Kipping, D.~M. 2013, Monthly Notices of the Royal Astronomical Society:
  Letters, 434, L51

\bibitem[{Klein {et~al.}(2021)Klein, Donati, H{\'e}brard, Zaire, Folsom, Morin,
  Delfosse, \& Bonfils}]{Klein2021}
Klein, B., Donati, J.-F., H{\'e}brard, {\'E}.~M., {et~al.} 2021, \mnras, 500,
  1844

\bibitem[{Kopparapu {et~al.}(2013)Kopparapu, Ramirez, Kasting, Eymet, Robinson,
  Mahadevan, Terrien, {Domagal-Goldman}, Meadows, \& Deshpande}]{Kopparapu2013}
Kopparapu, R.~K., Ramirez, R., Kasting, J.~F., {et~al.} 2013, \apj, 765, 131

\bibitem[{Kumaraswamy(1980)}]{Kumaraswamy1980}
Kumaraswamy, P. 1980, Journal of Hydrology, 46, 79

\bibitem[{Lafarga {et~al.}(2020)Lafarga, Ribas, Lovis, Perger, Zechmeister,
  Bauer, K{\"u}rster, {Cort{\'e}s-Contreras}, Morales, Herrero, Rosich, Baroch,
  Reiners, Caballero, Quirrenbach, Amado, Alacid, B{\'e}jar, Dreizler, Hatzes,
  Henning, Jeffers, Kaminski, Montes, Pedraz, {Rodr{\'i}guez-L{\'o}pez}, \&
  Schmitt}]{Lafarga2020}
Lafarga, M., Ribas, I., Lovis, C., {et~al.} 2020, \aap, 636, A36

\bibitem[{Laskar \& Petit(2017)}]{Laskar2017}
Laskar, J. \& Petit, A.~C. 2017, \aap, 605, A72

\bibitem[{{Lillo-Box} {et~al.}(2021){Lillo-Box}, Faria, Mascare{\~n}o,
  Figueira, Sousa, Tabernero, Lovis, Silva, Demangeon, Benatti, Santos, Mehner,
  Pepe, Sozzetti, Osorio, Hern{\'a}ndez, Micela, Hojjatpanah, Rebolo,
  Cristiani, Adibekyan, Allart, Prieto, Cabral, Damasso, Marcantonio, Curto,
  Martins, Megevand, Molaro, Nunes, Pall{\'e}, Pasquini, Poretti, \&
  Udry}]{Lillo-Box2021}
{Lillo-Box}, J., Faria, J.~P., Mascare{\~n}o, A.~S., {et~al.} 2021, \aap, 654,
  A60

\bibitem[{{Lillo-Box} {et~al.}(2020){Lillo-Box}, Figueira, Leleu, Acu{\~n}a,
  Faria, Hara, Santos, Correia, Robutel, Deleuil, Barrado, Sousa, Bonfils,
  Mousis, Almenara, {Astudillo-Defru}, Marcq, Udry, Lovis, \&
  Pepe}]{Lillo-Box2020}
{Lillo-Box}, J., Figueira, P., Leleu, A., {et~al.} 2020, \aap, 642, A121

\bibitem[{Lomb(1976)}]{Lomb1976}
Lomb, N.~R. 1976, Astrophysics and Space Science, 39, 447

\bibitem[{Mahadevan {et~al.}(2012)Mahadevan, Ramsey, Bender, Terrien, Wright,
  Halverson, Hearty, Nelson, Burton, Redman, Osterman, Diddams, Kasting, Endl,
  \& Deshpande}]{Mahadevan2012}
Mahadevan, S., Ramsey, L., Bender, C., {et~al.} 2012, 8446, 84461S

\bibitem[{Mann {et~al.}(2015)Mann, Feiden, Gaidos, Boyajian, \& {von
  Braun}}]{Mann2015}
Mann, A.~W., Feiden, G.~A., Gaidos, E., Boyajian, T., \& {von Braun}, K. 2015,
  \apj, 804, 64

\bibitem[{Mayor {et~al.}(2003)Mayor, Pepe, Queloz, Bouchy, Rupprecht, Lo~Curto,
  Avila, Benz, Bertaux, Bonfils, Dall, Dekker, Delabre, Eckert, Fleury,
  Gilliotte, Gojak, Guzman, Kohler, Lizon, Longinotti, Lovis, Megevand,
  Pasquini, Reyes, Sivan, Sosnowska, Soto, Udry, {van Kesteren}, Weber, \&
  Weilenmann}]{Mayor2003}
Mayor, M., Pepe, F., Queloz, D., {et~al.} 2003, The Messenger, 114, 20

\bibitem[{Meadows \& Barnes(2018)}]{Meadows2018}
Meadows, V.~S. \& Barnes, R.~K. 2018, in Handbook of {{Exoplanets}}, ed. H.~J.
  Deeg \& J.~A. Belmonte ({Cham}: {Springer International Publishing}),
  2771--2794

\bibitem[{Pavlenko {et~al.}(2017)Pavlenko, Su{\'a}rez~Mascare{\~n}o, Rebolo,
  Lodieu, B{\'e}jar, \& Gonz{\'a}lez~Hern{\'a}ndez}]{Pavlenko2017}
Pavlenko, Y., Su{\'a}rez~Mascare{\~n}o, A., Rebolo, R., {et~al.} 2017, \aap,
  606, A49

\bibitem[{Pepe {et~al.}(2021)Pepe, Cristiani, Rebolo, Santos, Dekker, Cabral,
  Di~Marcantonio, Figueira, Lo~Curto, Lovis, Mayor, M{\'e}gevand, Molaro, Riva,
  Zapatero~Osorio, Amate, Manescau, Pasquini, Zerbi, Adibekyan, Abreu,
  Affolter, Alibert, Aliverti, Allart, Allende~Prieto, {\'A}lvarez, Alves,
  Avila, Baldini, Bandy, Barros, Benz, Bianco, Borsa, Bourrier, Bouchy, Broeg,
  Calderone, Cirami, Coelho, Conconi, Coretti, Cumani, Cupani, D'Odorico,
  Damasso, Deiries, Delabre, Demangeon, Dumusque, Ehrenreich, Faria, Fragoso,
  Genolet, Genoni, G{\'e}nova~Santos, Gonz{\'a}lez~Hern{\'a}ndez, Hughes,
  Iwert, Kerber, Knudstrup, Landoni, Lavie, {Lillo-Box}, Lizon, Maire, Martins,
  Mehner, Micela, Modigliani, Monteiro, Monteiro, Moschetti, Murphy, Nunes,
  Oggioni, Oliveira, Oshagh, Pall{\'e}, Pariani, Poretti, Rasilla,
  Rebord{\~a}o, Redaelli, Santana~Tschudi, Santin, Santos, S{\'e}gransan,
  Schmidt, Segovia, Sosnowska, Sozzetti, Sousa, Span{\`o},
  Su{\'a}rez~Mascare{\~n}o, Tabernero, Tenegi, Udry, \& Zanutta}]{Pepe2021}
Pepe, F., Cristiani, S., Rebolo, R., {et~al.} 2021, \aap, 645, A96

\bibitem[{Perger {et~al.}(2021)Perger, {Anglada-Escud{\'e}}, Ribas, Rosich,
  Herrero, \& Morales}]{Perger2021}
Perger, M., {Anglada-Escud{\'e}}, G., Ribas, I., {et~al.} 2021, \aap, 645, A58

\bibitem[{Petit {et~al.}(2017)Petit, Laskar, \& Bou{\'e}}]{Petit2017}
Petit, A.~C., Laskar, J., \& Bou{\'e}, G. 2017, \aap, 607, A35

\bibitem[{Queloz {et~al.}(2009)Queloz, Bouchy, Moutou, Hatzes, H{\'e}brard,
  Alonso, Auvergne, Baglin, Barbieri, Barge, Benz, Bord{\'e}, Deeg, Deleuil,
  Dvorak, Erikson, Ferraz~Mello, Fridlund, Gandolfi, Gillon, Guenther, Guillot,
  Jorda, Hartmann, Lammer, L{\'e}ger, Llebaria, Lovis, Magain, Mayor, Mazeh,
  Ollivier, P{\"a}tzold, Pepe, Rauer, Rouan, Schneider, Segransan, Udry, \&
  Wuchterl}]{Queloz2009}
Queloz, D., Bouchy, F., Moutou, C., {et~al.} 2009, \aap, 506, 303

\bibitem[{Quirrenbach {et~al.}(2010)Quirrenbach, Amado, Mandel, Caballero,
  Mundt, Ribas, Reiners, Abril, Aceituno, Afonso, {Barrado y Navascues}, Bean,
  B{\'e}jar, Becerril, B{\"o}hm, C{\'a}rdenas, Claret, Colom{\'e}, Costillo,
  Dreizler, Fern{\'a}ndez, Francisco, Galad{\'i}, Garrido,
  Gonz{\'a}lez~Hern{\'a}ndez, Gu{\`a}rdia, Guenther, {Guti{\'e}rrez-Soto},
  Joergens, Hatzes, Helmling, Henning, Herrero, K{\"u}rster, Laun, Lenzen,
  Mall, Martin, {Mart{\'i}n-Ruiz}, Mirabet, Montes, Morales, Morales~Mu{\~n}oz,
  Moya, Naranjo, Rabaza, Ram{\'o}n, Rebolo, Reffert, Rodler, Rodr{\'i}guez,
  Rodr{\'i}guez~Trinidad, Rohloff, S{\'a}nchez~Carrasco, Schmidt, Seifert,
  Setiawan, Solano, Stahl, Storz, Su{\'a}rez, Thiele, Wagner, Wiedemann,
  Zapatero~Osorio, {del Burgo}, {S{\'a}nchez-Blanco}, \& Xu}]{Quirrenbach2010}
Quirrenbach, A., Amado, P.~J., Mandel, H., {et~al.} 2010, in {{SPIE
  Astronomical Telescopes}} + {{Instrumentation}}, ed. I.~S. McLean, S.~K.
  Ramsay, \& H.~Takami, {San Diego, California, USA}, 773513

\bibitem[{Rasmussen \& Williams(2006)}]{Rasmussen2006}
Rasmussen, C.~E. \& Williams, C. K.~I. 2006, Gaussian Processes for Machine
  Learning ({Cambridge}: {MIT Press})

\bibitem[{Ribas {et~al.}(2016)Ribas, Bolmont, Selsis, Reiners, Leconte,
  Raymond, Engle, Guinan, Morin, Turbet, Forget, \&
  {Anglada-Escud{\'e}}}]{Ribas2016}
Ribas, I., Bolmont, E., Selsis, F., {et~al.} 2016, \aap, 596, A111

\bibitem[{Ribas {et~al.}(2018)Ribas, Tuomi, Reiners, Butler, Morales, Perger,
  Dreizler, {Rodr{\'i}guez-L{\'o}pez}, Hern{\'a}ndez, Rosich, Feng, Trifonov,
  Vogt, Caballero, Hatzes, Herrero, Jeffers, Lafarga, Murgas, Nelson,
  Rodr{\'i}guez, Strachan, {Tal-Or}, Teske, {Toledo-Padr{\'o}n}, Zechmeister,
  Quirrenbach, Amado, Azzaro, B{\'e}jar, Barnes, Berdi{\~n}as, Burt, Coleman,
  {Cort{\'e}s-Contreras}, Crane, Engle, Guinan, Haswell, Henning, Holden,
  Jenkins, Jones, Kaminski, Kiraga, K{\"u}rster, Lee, {L{\'o}pez-Gonz{\'a}lez},
  Montes, Morin, Ofir, Pall{\'e}, Rebolo, Reffert, Schweitzer, Seifert,
  Shectman, Staab, Street, Mascare{\~n}o, Tsapras, Wang, \&
  {Anglada-Escud{\'e}}}]{Ribas2018}
Ribas, I., Tuomi, M., Reiners, A., {et~al.} 2018, \nat, 563, 365

\bibitem[{Ricker {et~al.}(2010)Ricker, Latham, Vanderspek, Ennico, Bakos,
  Brown, Burgasser, Charbonneau, Clampin, Deming, Doty, Dunham, Elliot, Holman,
  Ida, Jenkins, Jernigan, Kawai, Laughlin, Lissauer, Martel, Sasselov,
  Schingler, Seager, Torres, Udry, Villasenor, Winn, \& Worden}]{Ricker2010}
Ricker, G.~R., Latham, D.~W., Vanderspek, R.~K., {et~al.} 2010, American
  Astronomical Society Meeting Abstracts \#215, 450.06

\bibitem[{Santos {et~al.}(2014)Santos, Mortier, Faria, Dumusque, Adibekyan,
  {Delgado-Mena}, Figueira, Benamati, Boisse, Cunha, {Gomes da Silva},
  Lo~Curto, Lovis, Martins, Mayor, Melo, Oshagh, Pepe, Queloz, Santerne,
  S{\'e}gransan, Sozzetti, Sousa, \& Udry}]{Santos2014}
Santos, N.~C., Mortier, A., Faria, J.~P., {et~al.} 2014, \aap, 566, A35

\bibitem[{Scargle(1982)}]{Scargle1982}
Scargle, J.~D. 1982, \apj, 263, 835

\bibitem[{Seager {et~al.}(2010)Seager, Dotson, \& {Lunar and Planetary
  Institute}}]{Seager2010}
Seager, S., Dotson, R., \& {Lunar and Planetary Institute}, eds. 2010,
  Exoplanets, The {{University}} of {{Arizona}} Space Science Series ({Tucson :
  Houston}: {University of Arizona Press ; In collaboration with Lunar and
  Planetary Institute})

\bibitem[{Silva {et~al.}(2021)Silva, Faria, Santos, Sousa, Viana, Martins, \&
  Figueira}]{Silva2021}
Silva, A.~M., Faria, J.~P., Santos, N.~C., {et~al.} 2021, submitted to A\&A

\bibitem[{Su{\'a}rez~Mascare{\~n}o {et~al.}(2020)Su{\'a}rez~Mascare{\~n}o,
  Faria, Figueira, Lovis, Damasso, Gonz{\'a}lez~Hern{\'a}ndez, Rebolo,
  Cristiani, Pepe, Santos, Zapatero~Osorio, Adibekyan, Hojjatpanah, Sozzetti,
  Murgas, Abreu, Affolter, Alibert, Aliverti, Allart, Allende~Prieto, Alves,
  Amate, Avila, Baldini, Bandi, Barros, Bianco, Benz, Bouchy, Broeng, Cabral,
  Calderone, Cirami, Coelho, Conconi, Coretti, Cumani, Cupani, D'Odorico,
  Deiries, Delabre, Di~Marcantonio, Dumusque, Ehrenreich, Fragoso, Genolet,
  Genoni, G{\'e}nova~Santos, Hughes, Iwert, Kerber, Knusdstrup, Landoni, Lavie,
  {Lillo-Box}, Lizon, Lo~Curto, Maire, Manescau, Martins, M{\'e}gevand, Mehner,
  Micela, Modigliani, Molaro, Monteiro, Monteiro, Moschetti, Mueller, Nunes,
  Oggioni, Oliveira, Pall{\'e}, Pariani, Pasquini, Poretti, Rasilla, Redaelli,
  Riva, Santana~Tschudi, Santin, Santos, Segovia, Sosnowska, Sousa, Span{\`o},
  Tenegi, Udry, Zanutta, \& Zerbi}]{SuarezMascareno2020}
Su{\'a}rez~Mascare{\~n}o, A., Faria, J.~P., Figueira, P., {et~al.} 2020, \aap,
  639, A77

\bibitem[{Su{\'a}rez~Mascare{\~n}o {et~al.}(2016)Su{\'a}rez~Mascare{\~n}o,
  Rebolo, \& Gonz{\'a}lez~Hern{\'a}ndez}]{SuarezMascareno2016}
Su{\'a}rez~Mascare{\~n}o, A., Rebolo, R., \& Gonz{\'a}lez~Hern{\'a}ndez, J.~I.
  2016, \aap, 595, A12

\bibitem[{Su{\'a}rez~Mascare{\~n}o {et~al.}(2017)Su{\'a}rez~Mascare{\~n}o,
  Rebolo, Gonz{\'a}lez~Hern{\'a}ndez, \& Esposito}]{SuarezMascareno2017}
Su{\'a}rez~Mascare{\~n}o, A., Rebolo, R., Gonz{\'a}lez~Hern{\'a}ndez, J.~I., \&
  Esposito, M. 2017, \mnras, 468, 4772

\bibitem[{Tuomi {et~al.}(2019)Tuomi, Jones, Butler, Arriagada, Vogt, Burt,
  Laughlin, Holden, Shectman, Crane, Thompson, Keiser, Jenkins, Berdi{\~n}as,
  Diaz, Kiraga, \& Barnes}]{Tuomi2019}
Tuomi, M., Jones, H. R.~A., Butler, R.~P., {et~al.} 2019, arXiv:1906.04644
  [astro-ph] [\eprint[arXiv]{1906.04644}]

\bibitem[{Turbet {et~al.}(2016)Turbet, Leconte, Selsis, Bolmont, Forget, Ribas,
  Raymond, \& {Anglada-Escud{\'e}}}]{Turbet2016}
Turbet, M., Leconte, J., Selsis, F., {et~al.} 2016, \aap, 596, A112

\bibitem[{{Ulmer-Moll} {et~al.}(2019){Ulmer-Moll}, Santos, Figueira,
  Brinchmann, \& Faria}]{Ulmer-Moll2019}
{Ulmer-Moll}, S., Santos, N.~C., Figueira, P., Brinchmann, J., \& Faria, J.~P.
  2019, \aap, 630, A135

\bibitem[{Vida {et~al.}(2019)Vida, Ol{\'a}h, K{\H o}v{\'a}ri, {van
  Driel-Gesztelyi}, Mo{\'o}r, \& P{\'a}l}]{Vida2019}
Vida, K., Ol{\'a}h, K., K{\H o}v{\'a}ri, Z., {et~al.} 2019, ApJ, 884, 160

\bibitem[{Wargelin {et~al.}(2017)Wargelin, Saar, Pojma{\'n}ski, Drake, \&
  Kashyap}]{Wargelin2017}
Wargelin, B.~J., Saar, S.~H., Pojma{\'n}ski, G., Drake, J.~J., \& Kashyap,
  V.~L. 2017, \mnras, 464, 3281

\bibitem[{Wildi {et~al.}(2017)Wildi, Blind, Reshetov, Hernandez, Genolet,
  Conod, Sordet, Segovilla, Rasilla, Brousseau, Thibault, Delabre, Bandy,
  Sarajlic, Cabral, Bovay, Vall{\'e}e, Bouchy, Doyon, Artigau, Pepe, Hagelberg,
  Melo, Delfosse, Figueira, Santos, Hern{\'a}ndez, de~Medeiros, Rebolo, Broeg,
  Benz, Boisse, Malo, K{\"a}ufl, \& Saddlemyer}]{Wildi2017}
Wildi, F., Blind, N., Reshetov, V., {et~al.} 2017, in Techniques and
  {{Instrumentation}} for {{Detection}} of {{Exoplanets VIII}}, Vol. 10400
  ({International Society for Optics and Photonics}), 1040018

\bibitem[{Wildi {et~al.}(2010)Wildi, Pepe, Chazelas, Lo~Curto, \&
  Lovis}]{Wildi2010}
Wildi, F., Pepe, F., Chazelas, B., Lo~Curto, G., \& Lovis, C. 2010, in {{SPIE
  Astronomical Telescopes}} + {{Instrumentation}}, ed. I.~S. McLean, S.~K.
  Ramsay, \& H.~Takami, {San Diego, California, USA}, 77354X

\bibitem[{Zechmeister {et~al.}(2018)Zechmeister, Reiners, Amado, Azzaro, Bauer,
  B{\'e}jar, Caballero, Guenther, Hagen, Jeffers, Kaminski, K{\"u}rster,
  Launhardt, Montes, Morales, Quirrenbach, Reffert, Ribas, Seifert, {Tal-Or},
  \& Wolthoff}]{Zechmeister2018}
Zechmeister, M., Reiners, A., Amado, P.~J., {et~al.} 2018, \aap, 609, A12

\end{thebibliography}

\begin{appendix}

   \section{Analysis with different GP covariances}
   \label{app:covariances}

One important choice when modelling the stellar activity signal with a GP is the
kernel used to build the covariance matrix. Throughout the paper, we discussed
the results obtained with the QP kernel, which is given by

\begin{equation}
   \mathcal{K}_{\rm QP} (\tau) = 
         \eta_1^2 
         \exp\left[ - \frac{\tau^2}{2\eta_2^2}
                     - \frac{2\sin^2\left(\frac{\pi \, \tau}{\eta_3}\right)}{\eta_4^2} \right]
   \label{eq:qp-kernel}
,\end{equation}in terms of the time-lag $\tau=t_i - t_j$. This covariance is still the most
commonly used in the analysis of RV observations
\citep[e.g.][]{Grunblatt2015,Haywood2014}. Recently, \citet{Perger2021}
suggested a modification of the QP kernel to account for an explicit periodic
component at $\eta_3 / 2$, which is typically observed in simulated data. This
QPC kernel introduces one additional
hyperparameter, $\eta_5$, which controls the amplitude of the cosine term:

\begin{equation}
   \mathcal{K}_{\rm QPC} (\tau) = 
         \exp\left( - \frac{\tau^2}{2\eta_2^2} \right)
         \left[
            \eta_1^2 \exp\left(-\frac{2\sin^2\left(\frac{\pi \, \tau}{\eta_3}\right)}{\eta_4^2} \right)
            + \eta_5^2 \cos\left(4\pi\frac{\tau}{\eta_3}\right)
         \right]
         .
   \label{eq:qpc-kernel}
\end{equation}

\begin{figure}[h]
    \centering
    \includegraphics[width=\colfigsize]{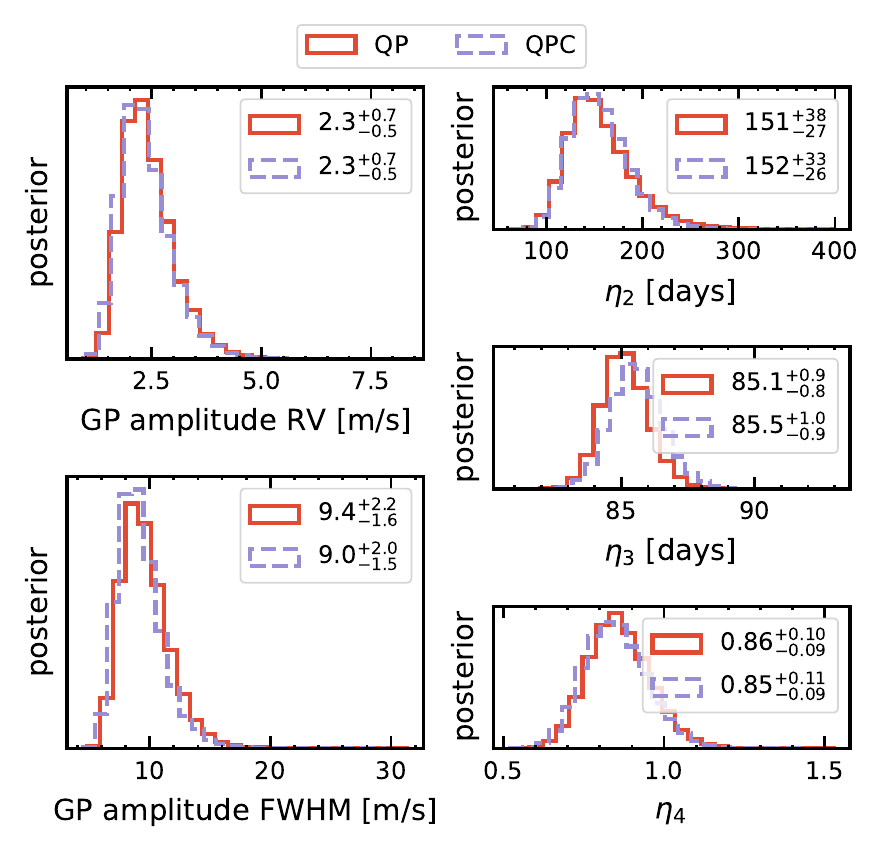}
    \caption{Posterior distributions for the hyperparameters of the GP kernels
    from the analysis of the TM RVs. We quote the posterior median and 68\%
    quantiles for each parameter and kernel.}
    \label{fig:gps_compare_etas}
\end{figure}

We performed the same analysis as in Sect. \ref{sec:rv-analysis}, replacing
the QP kernel with the QPC and assuming the same priors (see Appendix
\ref{app:priors}). We found virtually identical results to those obtained
with the QP kernel, as seen in \fig{fig:gps_compare_etas}, which compares the
posterior distributions for the kernel parameters.

Another assumption we made when jointly modelling RVs and FWHM was to share the
$\eta_2$, $\eta_3$, and $\eta_4$ hyperparameters between the two time series.
This is motivated by the expectation that the signal in RV and FWHM will share
the same correlation structure. However, stellar activity could appear in the
two observables with different characteristics. We nevertheless expect $\eta_3$
to correspond to the stellar rotation period and thus be the same in RV and
FWHM.
To test these assumptions, we repeated the analyses considering independent
$\eta_2$ and $\eta_4$ parameters for the RVs and the FWHM (with the same
priors). For both the CCF and TM RVs, this model revealed no significant
differences from the original model with shared parameters.

\subsection{Differential rotation and activity evolution}

\citet{Wargelin2017} found evidence for differential rotation (DR) on
\proxima by studying V-band observations from the ASAS survey. The stellar
rotation period changed between 77 and 90 days over a period of about 10
years, corresponding to a fractional DR estimate of $\Delta P_{\rm
rot}/\langle P_{\rm rot} \rangle \simeq 0.16$. 

Since the ESPRESSO dataset spans almost ten rotation periods and the three
subsets each span about one rotation period, it is interesting to check if
there are hints of DR showing in these data. While the GP model for stellar
activity does not explicitly account for DR, it can in
principle model its effects on both the RVs and the FWHM. 

\begin{figure}
    \centering
    \includegraphics[width=\colfigsize]{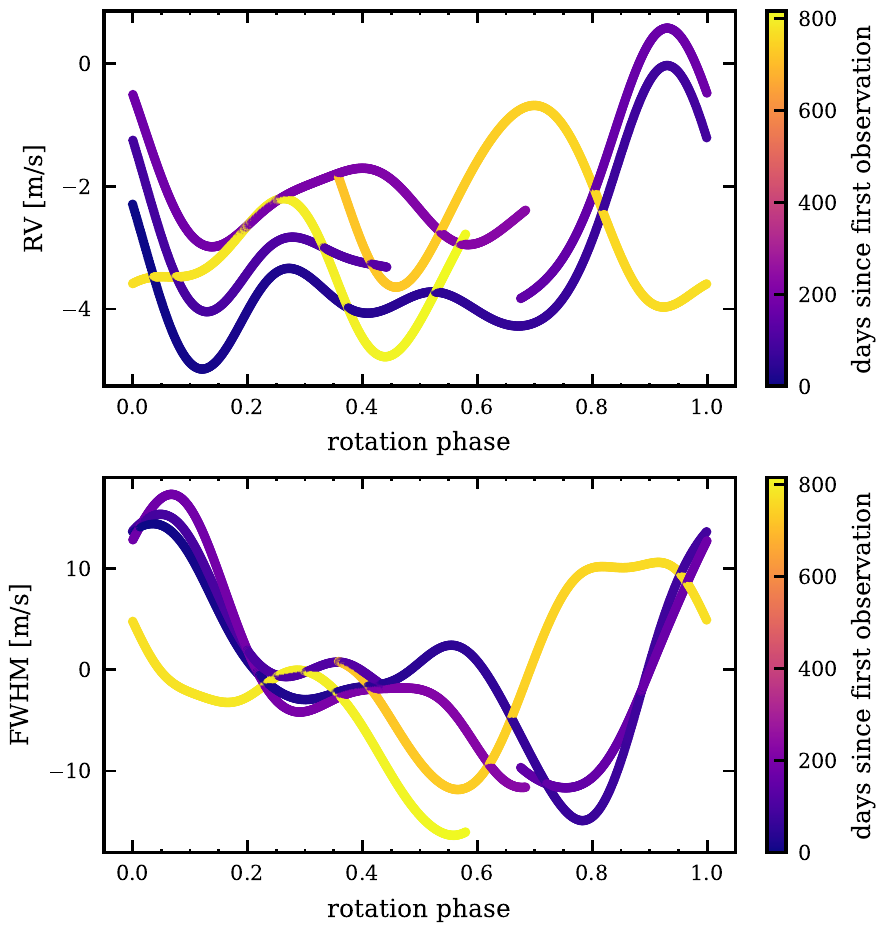}
    \caption{Predictive mean of the GP component of the model versus rotation
    phase, assuming a value of $\eta_3 = 85.1$ days. The top and bottom panels
    show the predictions for the RVs and the FWHM, respectively. The curves are
    only shown for times in which there are ESPRESSO observations, and their
    colour reflects the time since the first observation.}
    \label{fig:phased_gp}
\end{figure}

We use the maximum likelihood solution from the model with two planets fit to
the TM RVs. Figure \ref{fig:phased_gp} shows the predictive mean for the GP
component (after subtracting the two Keplerian signals and the quadratic trend),
phase-folded on the value of $\eta_3$, with the colour of the curves
corresponding to the time since the first observation. The predictive is only
shown for those times in which there are observations (that is, excluding the
gap between ESPRESSO19 and ESPRESSO21) because, in the absence of data, the GP
tends to the mean of zero and its interpretation is less meaningful.

At a glance, \fig{fig:phased_gp} highlights the quasi-periodicity of the
activity signal, as modelled by the GP. Moreover, as is also visible in
\fig{fig:2p_solution} below, there is a clear time delay between the activity
signal on the RVs and on the FWHM, with the maxima (and minima) of the FWHM
happening 10-15 days, or about 15\% of the rotation period, after the RV maxima.
This temporal shift has been observed before for the Sun
\citep{CollierCameron2019} and other stars \citep[e.g.][]{Queloz2009,Santos2014}
and is related to the effect of active regions on the stellar line profile
\citep[see][and references therein]{CollierCameron2019}.

Perhaps more interestingly, \fig{fig:phased_gp} also shows a small variation in
the rotation period, as measured by the position of the maxima and minima of the
activity signal, over short timescales of two to three rotation periods (c.f.
the blue and purple minima seen on the left in the top and bottom panels).
Assuming that the same active regions are creating the signal over these
timescales, such a variation could be due to DR as the active regions move
differentially on the stellar surface.

   \section{Prior distributions}
   \label{app:priors}

   Here we detail the prior distributions and further justify some of the
   choices made when defining them. For clarity, we refer to three distinct
   components of the model: background, Keplerians, and noise. 
   
   The background model includes the systemic velocity and the quadratic trend
   coefficients in the RVs, as well as the `systemic FWHM', which is a simple
   mean value for this quantity. Our priors for $v_{\rm sys}$ and $f_{\rm sys}$
   are slightly unusual, but simply limit the parameters to the observed ranges
   and do not influence the results, especially when taking into account that
   the orbital period of any Keplerian component is limited to the time span of
   the data (see below). For the RV and FWHM offsets between the (three) subsets
   of ESPRESSO data, we chose uniform priors within the observed ranges. The
   Gaussian priors for the quadratic trend coefficients are, again, informed by
   the data but only to the extent of setting typical scales for these
   parameters.

   Our rationale when setting priors for the orbital parameters was to use the
   same uninformative distributions for all $N_p$ Keplerians included in the
   model. In principle, there is prior information about \proxima b that is
   independent of the ESPRESSO data, and so could be used. In practice, the
   signal is so clearly detected in the ESPRESSO data that this prior becomes
   less relevant. On the other hand, there is no prior information about the
   5-day signal that can be used since its detection relies solely on ESPRESSO
   data. A log-uniform prior limited at the time span of the data is thus a
   reasonable choice.
   
   The most restrictive prior ends up being the one for $K$, which we limit to
   10 \ms. This is simply because there is no indication of higher-amplitude
   signals being present in the ESPRESSO data, and a prior that used a simple
   statistic of the observed RVs (e.g. the RV span) would be virtually the
   same. More than 99.9\% of the posterior probability ends up being below 1.6
   \ms, suggesting that this choice of upper limit does not influence the
   results. The Kumaraswamy prior for the orbital eccentricities and its
   parameters are motivated by \citet{Kipping2013} where a Beta distribution was
   proposed. The Kumaraswamy is virtually identical to the Beta
   \citep{Kumaraswamy1980}, but its cumulative distribution function (required
   by DNS) is analytical.

   In the noise component, we include the GP and the individual jitters for each
   subset of data. The hyperparameters of the QP kernel are $\eta_{1-4}$ (see
   Eq. \ref{eq:qp-kernel}) and we consider independent $\eta_1$ parameters for
   the RVs and the FWHM. These parameters are assigned modified log-uniform
   priors up to the observed range and which extend to zero. The justification
   for the prior for $\eta_3$ was mentioned before, with the wide uniform prior
   we use being quite conservative. Since we expect the activity signal to be
   QP, the prior for $\eta_2$ starts at the lower limit for
   $\eta_3$, and extends to a reasonable limit of 400 days, almost five times
   the rotation period of \proxima. We chose a log-uniform prior due to the
   large span. We also tested a different prior for $\eta_2$ that extended down
   to zero (an inverse Gamma distribution) but this did not change the results.

   \begin{table}[h!]
      \renewcommand{\arraystretch}{1.15}
      \caption{Parameters and prior distributions of our model for the RV and
      FWHM data.}
      \label{tab:priors}
      \centering
      \begin{tabular}{l l l}
        \hline\hline
        Parameter & Units & Prior \\
        \hline
        $v_{\rm sys}$ & \ms & $\mathcal{U}$ ($\min$ RV, $\max$ RV) \\
        $f_{\rm sys}$ & \ms & $\mathcal{U}$ ($\min$ FWHM, $\max$ FWHM) \\
        slope & \ms/day     & $\mathcal{G} \left(0, 10^{\,\Delta {\rm RV} / {\Delta t}}\right)$   \\
        quadr & \ms/day$^2$ & $\mathcal{G} \left(0, 10^{\,\Delta {\rm RV} / {\Delta t^2}}\right)$ \\[0.5em]
        $j^{\,\rm RV}$ & \ms & $\mathcal{M}\mathcal{L}\mathcal{U}$ (0.1\,$\Delta$RV, $\Delta$RV) \\
        $j^{\,\rm FWHM}$ & \ms & $\mathcal{M}\mathcal{L}\mathcal{U}$ (1, $\Delta$FWHM) \\[0.5em]
        $P$ & days & $\mathcal{L}\mathcal{U}$ (1, $\Delta t$) \\
        $K$ & \ms & $\mathcal{M}\mathcal{L}\mathcal{U}$ (0, 10) \\
        $e$ &  &  $\mathcal{K}(0.867, 3.03)$ \\
        $M_0$ &  & $\mathcal{U}$ (0, $2\pi$) \\
        $\omega$ &  & $\mathcal{U}$ (0, $2\pi$) \\[0.5em]
        $\eta_1$ RV & \ms & $\mathcal{M}\mathcal{L}\mathcal{U}$ (1, $\Delta$RV) \\
        $\eta_1$ FWHM & \ms & $\mathcal{M}\mathcal{L}\mathcal{U}$ (1, $\Delta$FWHM) \\
        $\eta_2$ & days & $\mathcal{L}\mathcal{U}$ (60, 400) \\
        $\eta_3$ & days & $\mathcal{U}$ (60, 100) \\
        $\eta_4$ &  & $\mathcal{L}\mathcal{U}$ (0.1, 10) \\[0.2em]
        $\eta_{5,\rm QPC}$ RV & \ms & $\mathcal{M}\mathcal{L}\mathcal{U}$ (1, $\Delta$RV) \\
        $\eta_{5,\rm QPC}$ FWHM & \ms & $\mathcal{M}\mathcal{L}\mathcal{U}$ (1, $\Delta$FWHM) \\
        \hline
      \end{tabular}
   \end{table}

   \section{Posterior estimates}
   \label{app:posterior_estimates}

\begin{figure*}
   \centering
   \includegraphics[width=\hsize]{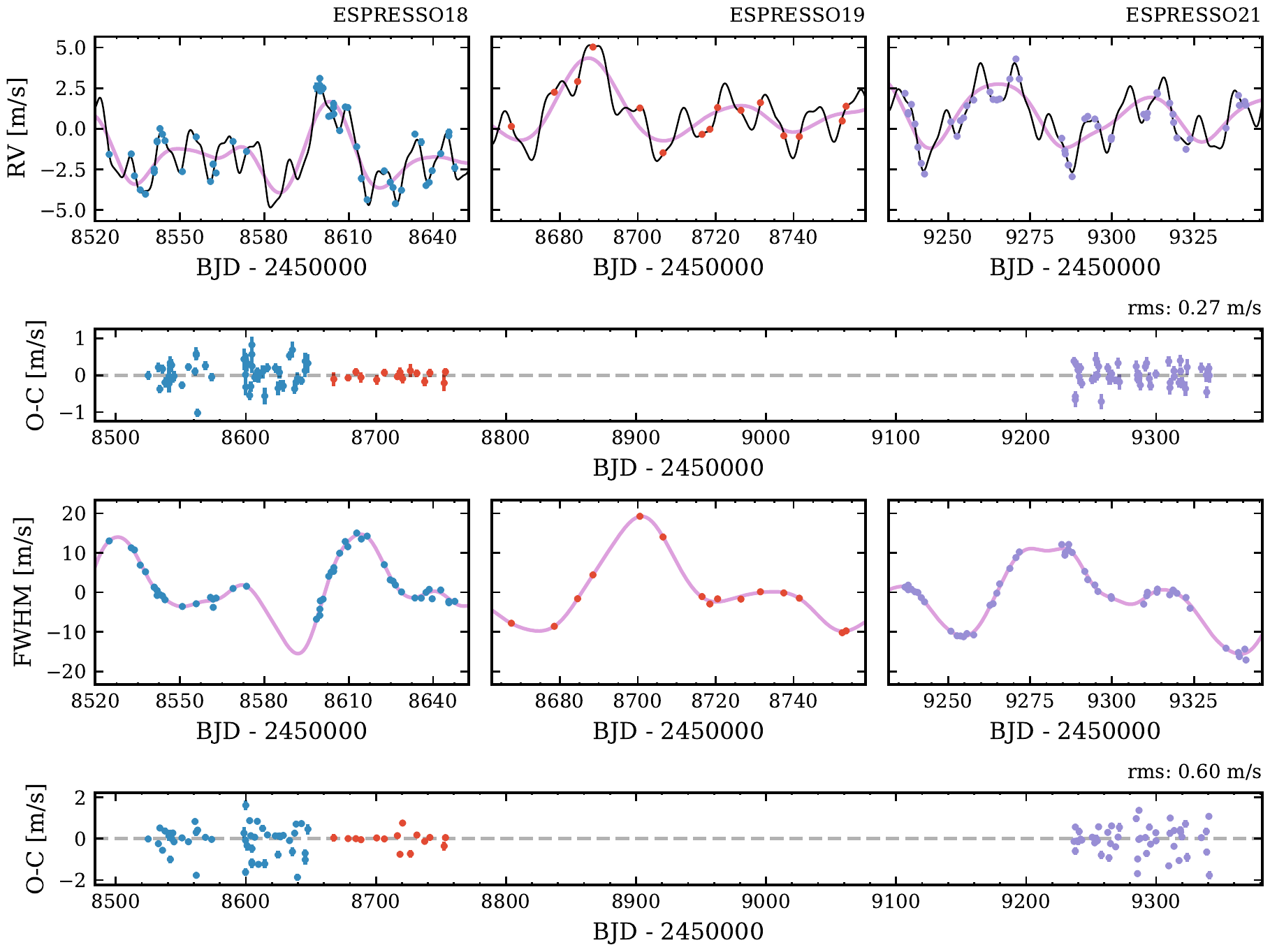}
   \caption{Maximum \emph{a posteriori} solution for the two-planet model on the
      TM RVs. The top panels show the RV observations for ESPRESSO18,
      ESPRESSO19, and ESPRESSO21, together with the GP component of the model
      (in pink) as well as the full model (in black). The RV residuals are shown
      just below, highlighting the full residual rms. The two bottom panels show
      the same for the FWHM.}
   \label{fig:2p_solution}
\end{figure*}

\begin{figure*}
   \centering
   \includegraphics[width=0.49\hsize]{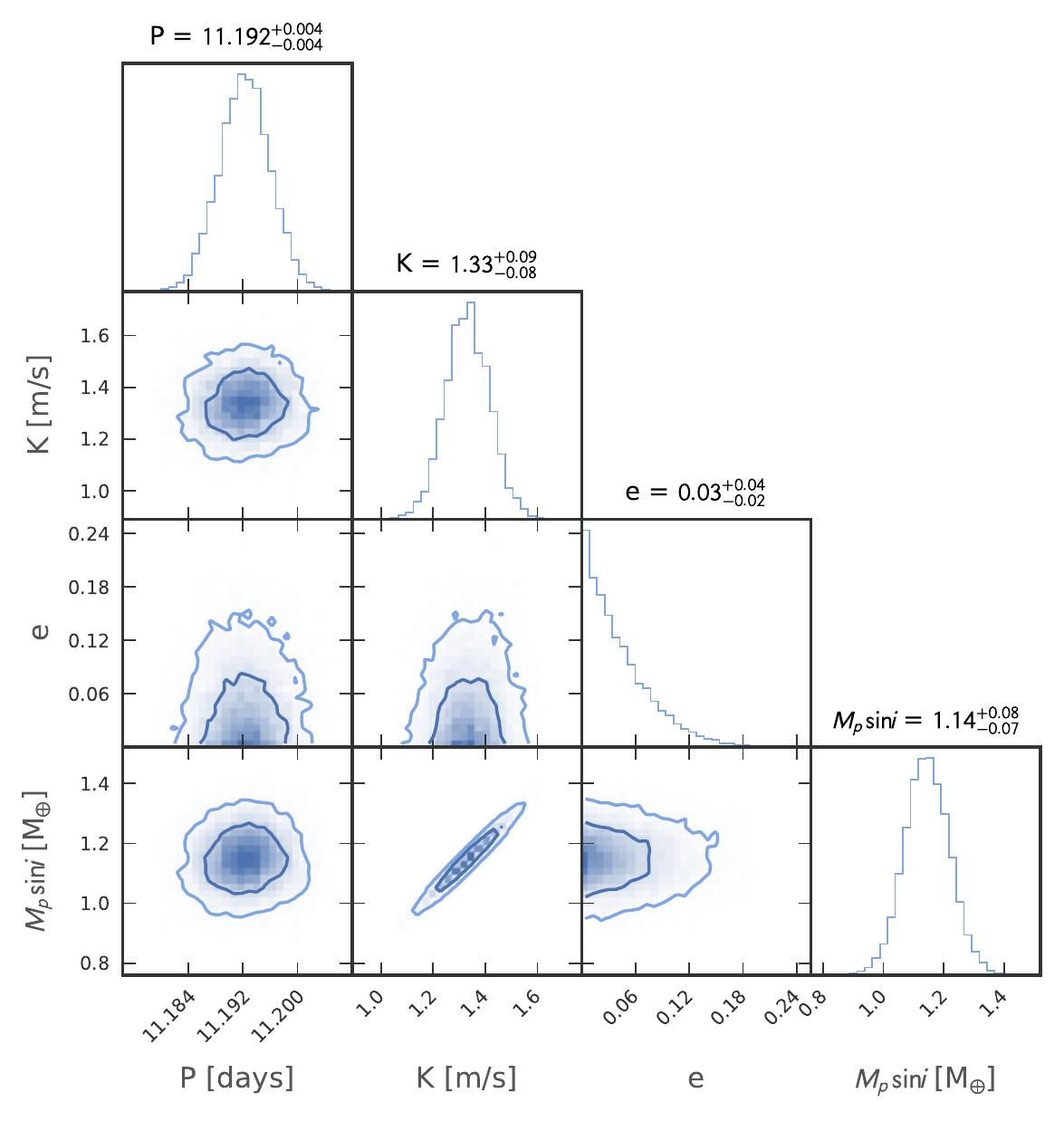}
   \includegraphics[width=0.49\hsize]{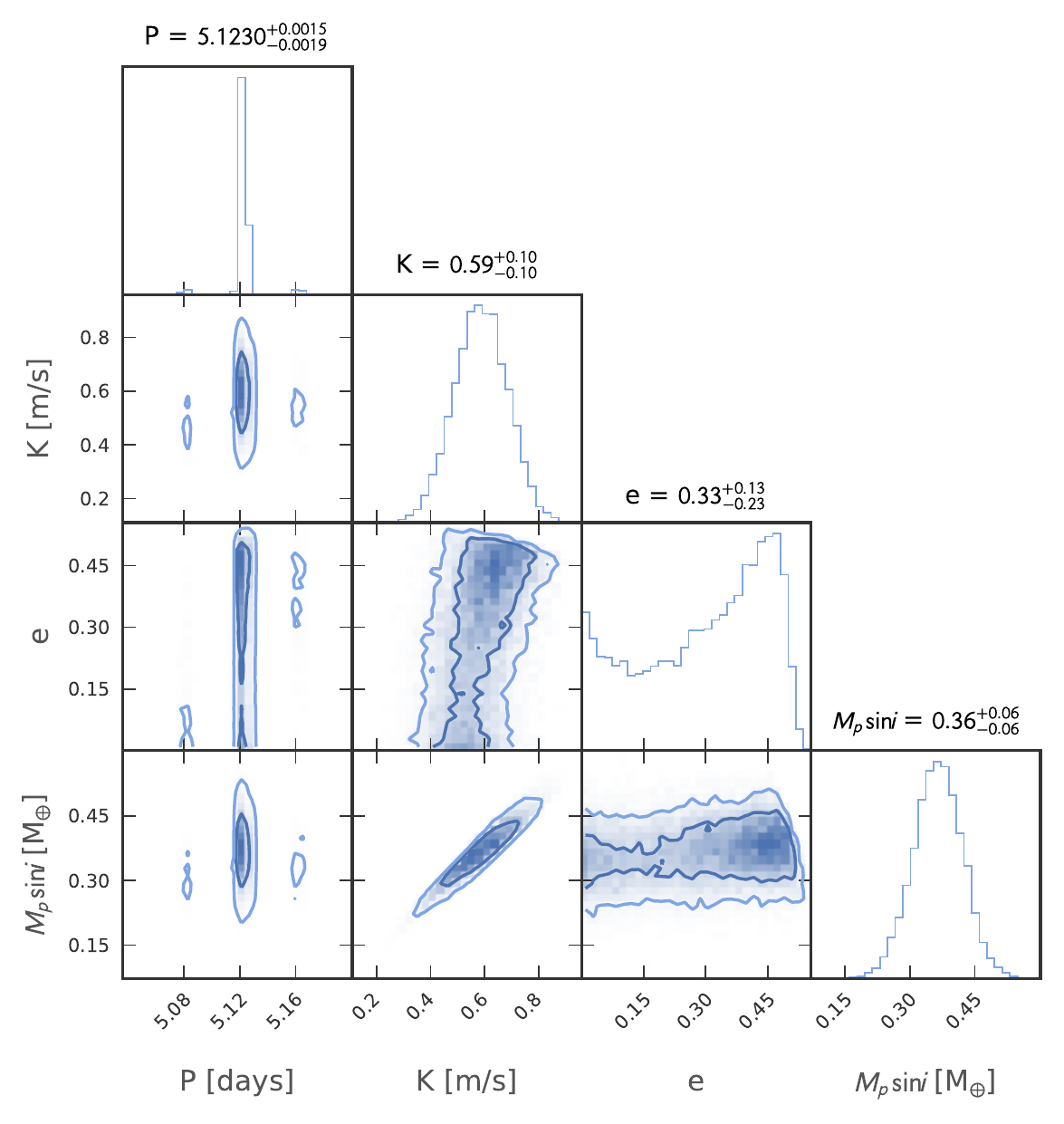}
   \caption{Joint and marginal posteriors for the orbital periods,
         semi-amplitudes, eccentricities, and minimum planet masses,
         $M_p\!\sin\!i$, in Earth masses, of the 11-day (left) and 5-day
         (right) Keplerian signals. These results come from the analysis of
         the CCF RVs. The estimates at the top of each panel correspond to the
         median and 68\% quantiles of the posteriors.}
   \label{fig:corners_planets_CCF}
\end{figure*}

Table \ref{tab:post} lists the estimates for the parameters of the two-planet
model resulting from the analysis of the CCF and TM RVs. Each estimate is shown
as the (MAP) value of the parameter together with the 68\% quantiles of the
posterior distribution.
In \fig{fig:2p_solution}, we show the fit to the ESPRESSO TM RVs and FWHM,
individually for each subset of the data together with the residuals after
subtracting the complete model. These panels also show the weighted RMS of the
residuals assuming the original RV and FWHM uncertainties (i.e. the jitter
parameters have not been added in quadrature).

Finally, \fig{fig:corners_planets_CCF} shows the joint posteriors for the
orbital period, semi-amplitude, eccentricity, and planet minimum mass of the two
Keplerians from the analysis of the CCF RVs.  In this dataset, the
semi-amplitude of the 5-day signal is significantly smaller than for the CCF
RVs, at $38$ \cms (\fig{fig:corners_planets_TM}), although the results for the
two datasets are compatible at less than 2$\sigma$. This ultimately leads to an
estimate for the planetary mass that is 30\% lower than with the CCF RVs, at
$0.24\pm0.05$ \Mearth. The eccentricity is better constrained by the TM RVs and
the solutions for both planets are compatible with circular orbits.

\begin{table*}
    \renewcommand{\arraystretch}{1.3}
      \caption{Posterior estimates for the parameters of the two-planet model from
      the analysis of CCF and TM RVs. For each parameter we show the MAP
      estimate and the 68\% quantiles of the distribution.}
      \label{tab:post}
      \centering
      \begin{tabular}{l c c c}
        \hline\hline
        \multirow{3}{*}{Parameter} & \multicolumn{2}{c}{$N_p=2$ model} & \multirow{3}{*}{Units} \\ 
                                   & CCF RVs & TM RVs                  &                            \\ 
                                   &         & (adopted)               &                            \\ 
        \hline
        \multicolumn{1}{c}{Keplerians}                                                                                 \\
        \multicolumn{1}{c}{planet \emph{b}}                                                                            \\
  $P$                            & $11.193 ^{+0.004} _{-0.004}$      & $11.1868 ^{+0.0029} _{-0.0031}$   & days       \\
  $K$                            & $1.33 ^{+0.09} _{-0.08}$          & $1.24 ^{+0.07} _{-0.07}$          & \ms        \\
  $M_0$                          & $0.8 ^{+2.3} _{-1.9}$             & $5.0 ^{+1.9} _{-2.1}$             &            \\
  $e$                            & $0.002 ^{+0.046} _{-0.002}$       & $0.02 ^{+0.04} _{-0.02}$          &            \\
  $\omega$                       & $1.3 ^{+2.3} _{-2.3}$             & $3.3 ^{+1.8} _{-2.3}$             &            \\
  &                                   &                                   &            \\
  $M_p \,\sin\!i$                & $1.15 ^{+0.08} _{-0.07}$          & $1.07 ^{+0.06} _{-0.06}$          & $M_\oplus$ \\
  $a$                            & $0.04858 ^{+0.00029} _{-0.00029}$ & $0.04856 ^{+0.00030} _{-0.00030}$ & au         \\
        \multicolumn{1}{c}{planet \emph{d}}                                                                            \\
  $P$                            & $5.1243 ^{+0.0015} _{-0.0019}$    & $5.122 ^{+0.002} _{-0.036}$       & days       \\
  $K$                            & $0.65 ^{+0.10} _{-0.10}$          & $0.39 ^{+0.07} _{-0.07}$          & \ms        \\
  $M_0$                          & $2.2 ^{+0.4} _{-0.5}$             & $5.5 ^{+2.8} _{-1.6}$             &            \\
  $e$                            & $0.37 ^{+0.13} _{-0.23}$          & $0.04 ^{+0.15} _{-0.04}$          &            \\
  $\omega$                       & $1.3 ^{+0.8} _{-0.4}$             & $4.0 ^{+2.0} _{-1.7}$             &            \\
  $M_p \,\sin\!i$                & $0.40 ^{+0.06} _{-0.06}$          & $0.26 ^{+0.05} _{-0.05}$          & $M_\oplus$ \\
  $a$                            & $0.02886 ^{+0.00018} _{-0.00018}$ & $0.02885 ^{+0.00019} _{-0.00022}$ & au         \\
   \multicolumn{1}{c}{GP}                                                                                              \\
   $\eta_1$ RV                    & $1.3 ^{+0.6} _{-0.4}$             & $1.7 ^{+0.7} _{-0.5}$             & \ms        \\
   $\eta_1$ FWHM                  & $7.5 ^{+2.2} _{-1.6}$             & $6.6 ^{+2.2} _{-1.6}$             & \ms        \\
   $\eta_2$                       & $142 ^{+42} _{-30}$               & $156 ^{+38} _{-27}$               & days       \\
   $\eta_3$                       & $84.5 ^{+1.1} _{-0.9}$            & $84.5 ^{+0.9} _{-0.8}$            & days       \\
   $\eta_4$                       & $0.71 ^{+0.11} _{-0.09}$          & $0.72 ^{+0.10} _{-0.09}$          &            \\
   \multicolumn{1}{c}{noise}                                                                                           \\
   $j^{RV}_{\rm ESP18}$           & $0.33 ^{+0.09} _{-0.08}$          & $0.40 ^{+0.07} _{-0.05}$          & \ms        \\
   $j^{RV}_{\rm ESP19}$           & $0.07 ^{+0.25} _{-0.07}$          & $0.07 ^{+0.19} _{-0.07}$          & \ms        \\
   $j^{RV}_{\rm ESP21}$           & $0.39 ^{+0.08} _{-0.07}$          & $0.26 ^{+0.05} _{-0.05}$          & \ms        \\
   $j^{FWHM}_{\rm ESP18}$         & $0.65 ^{+0.18} _{-0.16}$          & $0.56 ^{+0.17} _{-0.16}$          & \ms        \\
   $j^{FWHM}_{\rm ESP19}$         & $0.3 ^{+0.6} _{-0.3}$             & $0.03 ^{+0.54} _{-0.03}$          & \ms        \\
   $j^{FWHM}_{\rm ESP21}$         & $0.55 ^{+0.14} _{-0.13}$          & $0.60 ^{+0.14} _{-0.13}$          & \ms        \\
   \multicolumn{1}{c}{background}                                                                                      \\
   RV slope                       & $-1.0 ^{+1.8} _{-2.0}$            & $-1.7 ^{+2.4} _{-2.1}$            & \msy       \\
   RV quadr                       & $0.5 ^{+1.3} _{-1.3}$             & $0.0 ^{+1.3} _{-1.3}$             & \msyy      \\
   RV offset ESP18-ESP21          & $-4 ^{+3} _{-4}$                  & $-5 ^{+5} _{-4}$                  & \ms        \\
   RV offset ESP19-ESP21          & $-1 ^{+3} _{-4}$                  & $-2 ^{+4} _{-4}$                  & \ms        \\
   FWHM offset ESP18-ESP21        & $-2 ^{+8} _{-8}$                  & $2 ^{+8} _{-8}$                   & \ms        \\
   FWHM offset ESP19-ESP21        & $1 ^{+8} _{-8}$                   & $3 ^{+8} _{-8}$                   & \ms        \\
   $v_{\rm sys}$                  & $-21361.9 ^{+2.3} _{-2.1}$        & $-21357.1 ^{+2.4} _{-2.8}$        & \ms        \\
   $f_{\rm sys}$                  & $3845 ^{+6} _{-6}$                & $3844 ^{+6} _{-6}$                & \ms        \\
  \end{tabular}
    \tablefoot{Planet minimum masses and semi-major axes were derived assuming
               the stellar mass given in Table \ref{tab:parameters}.}
\end{table*}

\end{appendix}

\end{document}